\documentclass[11pt,uplatex]{article}
\usepackage{graphicx}
\usepackage{color}
\usepackage{amsmath, amssymb, amsthm, ascmac, comment, enumerate, bm, newtxtext}
\usepackage{booktabs, latexsym, mathabx, mathrsfs, mathtools, psfrag, setspace, algorithm, algpseudocode, subcaption}
\usepackage{natbib}
\usepackage{chngcntr}

\counterwithin{figure}{section}
\counterwithin{algorithm}{section}

\mathtoolsset{showonlyrefs=true}

\definecolor{mycolor}{cmyk}{0.82, 0.23, 0.26, 0}
\usepackage[colorlinks=true, linkcolor=mycolor, filecolor=mycolor, urlcolor=mycolor, citecolor=mycolor, driverfallback=dvipdfmx]{hyperref}
\usepackage[margin = 2.4cm]{geometry}

\usepackage{titlesec}
\titleformat*{\section}{\Large\bfseries\sffamily}
\titleformat*{\subsection}{\large\bfseries\sffamily}
\titleformat*{\subsubsection}{\large\bfseries\sffamily}

\DeclareMathOperator*{\argmin}{argmin}
\DeclareMathOperator*{\argmax}{argmax}

\newcommand{\indep}{\mathrel{\text{\scalebox{1.07}{$\perp\mkern-10mu\perp$}}}}
\renewcommand{\hat}{\widehat}
\renewcommand{\tilde}{\widetilde}

\renewcommand{\bar}{\overline}
\newcommand{\bbE}{\mathbb{E}}

\newcommand{\bbR}{\mathbb{R}}
\newcommand{\bbB}{\mathbb{B}}
\newcommand{\bbG}{\mathbb{G}}
\newcommand{\bbK}{\mathbb{K}}
\newcommand{\calB}{\mathcal{B}}
\newcommand{\calC}{\mathcal{C}}
\newcommand{\calE}{\mathcal{E}}
\newcommand{\calG}{\mathcal{G}}
\newcommand{\calI}{\mathcal{I}}
\newcommand{\calJ}{\mathcal{J}}
\newcommand{\calS}{\mathcal{S}}
\newcommand{\calX}{\mathcal{X}}
\newcommand{\calP}{\mathcal{P}}
\newcommand{\calT}{\mathcal{T}}

\numberwithin{equation}{section}
 
\allowdisplaybreaks[2]

\theoremstyle{definition}
\newtheorem{theorem}{Theorem}[section]
\newtheorem{assumption}{Assumption}[section]
\newtheorem{definition}{Definition}[section]

\newtheorem{example}{Example}[section]

\newtheorem{proposition}{Proposition}[section]
\newtheorem{remark}{Remark}[section]

\title{Evaluating Policy Effects under Network Interference without Network Information: A Transfer Learning Approach}

\author{Tadao Hoshino\thanks{School of Political Science and Economics, Waseda University, 1-6-1 Nishi-waseda, Shinjuku-ku, Tokyo 169-8050, Japan. Email: \href{mailto:thoshino@waseda.jp}{thoshino@waseda.jp}.}}

\begin{document}
\maketitle

\begin{abstract}    
    This paper develops a sensitivity analysis framework that transfers the average total treatment effect (ATTE) from source data with a fully observed network to target data whose network is completely unknown.
    The ATTE represents the average social impact of a policy that assigns the treatment to every individual in the dataset.
    We postulate a covariate-shift type assumption that both source and target datasets share the same conditional mean outcome.
    However, because the target network is unobserved, this assumption alone is not sufficient to pin down the ATTE for the target data.
    To address this issue, we consider a sensitivity analysis based on the uncertainty of the target network's degree distribution, where the extent of uncertainty is measured by the Wasserstein distance from a given reference degree distribution.
    We then construct bounds on the target ATTE using a linear programming-based estimator.
    The limiting distribution of the bound estimator is derived via the functional delta method, and we develop a wild bootstrap approach to approximate the distribution.
    As an empirical illustration, we revisit the social network experiment on farmers' weather insurance adoption in China by Cai et al. (2015).
\end{abstract}

\clearpage

\section{Introduction}

Randomized controlled trials (RCTs) have long been the gold standard for estimating causal effects.  
However, it is rare that the group of individuals of interest for whom researchers or policymakers wish to know causal effects precisely coincides with the experimental sample.  
In many cases, the purpose of conducting an RCT is to determine in advance whether a treatment of concern yields positive effects so that it can then be introduced to a target population of real interest.  

Nevertheless, the causal effects estimated from the experimental data cannot, in general, be directly applied to the non-experimental target data.
To transfer estimation results from the source to the target data, we need to employ some data-adaptation techniques -- \textit{causal transfer learning}, transfer learning methods to infer causal effects in the target data, optimal treatment rules, and so forth.  
There is a rapidly growing body of literature developing transfer learning methods in this context (e.g., \citealp{stuart2011use, hartman2015sample, buchanan2018generalizing, wu2023transfer}, among many others).  
For comprehensive surveys and tutorials, see, for example, \cite{dahabreh2020extending} and \cite{degtiar2023review}.  

Meanwhile, causal inference under network interference has gained increasing attention in the literature across economics, education, epidemiology, political science, and related areas.
In these literature, performing an RCT has become one of major approaches for estimating treatment effects and \textit{spillover effects} -- the effects of others' treatments on one's own outcome (e.g., \citealp{bond201261, cai2015social, paluck2016changing, carter2021subsidies}, among many others).
While these studies have revealed both own and spillover effects in their experimental samples to some extent, policymakers ultimately may wish to extrapolate such findings to larger populations of their real concern.
However, to the best of our knowledge, in contrast to the rich body of studies without network interactions, there are few, if any, studies that explicitly consider the transferability of causal effects under network interference.

The purpose of this paper is to fill this gap.
Specifically, we propose a framework for inferring causal policy effects in target network data by transferring results obtained from source network data.
In particular, we focus on the situation in which only individual covariates (or their distributions) are available for the target data but its network structure is completely unknown.
Such situations are typical.
For example, when evaluating infection prevention policies such as mandatory face-mask wearing or vaccination, the target population of interest for policymakers is the entire country.
Collecting detailed network information for all citizens would be prohibitively costly, whereas demographic variables are often readily available from surveys and the census.
As another example, suppose a financial company wishes to promote its insurance or savings products for farmers.
Using an RCT among Chinese rice farmers, \cite{cai2015social} show that holding detailed information sessions significantly increases insurance take-up through social networks in each village.
Given this evidence, the insurer might wish to scale up the same sessions nationwide.
In that case, the target population is all farmers in the country, but information on the social networks in all villages is usually unavailable.

\bigskip

In order to transfer results from one sample to another, it is generally necessary to impose some similarity (or transferability) condition that links the two samples.
A common condition of this kind is the so-called \textit{covariate shift}, which assumes that the two datasets share common conditional mean potential outcome functions, while the covariate distributions may differ.
When the objective is merely to estimate the conditional mean potential outcome, as is often the case in the causal inference literature, the covariate-shift assumption alone suffices.
However, from a policymaker's perspective, the goal is often to assess the expected social impact of a specific policy, rather than to estimate the conditional mean function itself.
Motivated by this, we focus on the policy that assigns treatment to every unit in the dataset.
Then, the causal parameter of interest in this context is the average total treatment effect (ATTE) over the target data.
The total treatment effect is defined as the difference in potential outcomes when all units are assigned to treatment versus when all units are assigned to control; this is also referred to as the global treatment effect (e.g., \citealp{chin2019regression, ugander2023randomized, faridani2024linear}).
In policy settings such as nationwide infection-prevention campaigns or the promotion of insurance services to all farmers, as in the examples above, the ATTE should be a natural target parameter.

In the absence of network information in the target data, the covariate-shift assumption alone is not sufficient to point estimate the target ATTE.
To address this issue, we propose to conduct a sensitivity analysis with respect to the target network's degree distribution.
Specifically, following the idea of Wasserstein distributionally robust optimization (e.g., \citealp{blanchet2019quantifying, blanchet2021statistical, gao2023distributionally}), we quantify the uncertainty in the target degree distribution using the Wasserstein distance from a given reference distribution.
While in the literature of sensitivity analysis on distributional uncertainty, the Kullback-Leibler divergence is more commonly used (e.g., \citealp{duchi2021learning, spini2021robustness, christensen2023counterfactual}), the Wasserstein distance offers several practical merits, such as allowing non-overlapping supports and  computational simplicity.
We show that the resulting bound estimator for the target ATTE can be obtained by solving a set of simple linear programming problems.
Under regularity conditions, we derive the limiting distribution of the bound at each Wasserstein radius via the functional delta method (\citealp{fang2019inference}).
Moreover, we propose a dependent wild bootstrap method to approximate the distribution of the bound, following the approach of \cite{conley2023bootstrap}.

As an empirical illustration, we apply our method to data from a field experiment conducted by \cite{cai2015social}, which investigated how social networks affect the adoption of a weather insurance product among rural Chinese rice farmers.
To evaluate the performance of our sensitivity analysis framework, we randomly partition the villages into source and target groups and estimate the bound on the ATTE for the target group.
Since the full network information is available for all villages, we are able to compute the point estimate of the target ATTE and examine how the choice of the Wasserstein radius influences the coverage of the ATTE.
In this setting, since both groups are essentially drawn from a common population (i.e., they are all rice farmers in the same Chinese province), the resulting bounds are shown to be very informative even under small Wasserstein radius.

\paragraph{Paper organization}

The remainder of the paper is organized as follows.
Section \ref{sec:setup} formally presents our problem setup and the parameter of primary interest, the ATTE.
Section \ref{sec:LP} provides a linear-programming characterization of the Wasserstein bound on the target ATTE.
The estimation of the bound and its asymptotic properties are discussed in Section \ref{sec:estimation}, where we also introduce a wild bootstrap procedure for inference.
In Section \ref{sec:MC}, we conduct a set of Monte Carlo simulations to examine the finite sample performance of the proposed inference method.
Section \ref{sec:empiric} presents an empirical illustration based on the data from \cite{cai2015social}.
Section \ref{sec:conclusion} concludes.
The appendix contains proofs and additional technical supplementary material.

\section{Problem Setup}\label{sec:setup}

Consider a set of observations $\calI$ of size $n^\calI = |\calI|$, which we label the experimental sample.
Here, "experimental" is for terminology purposes only, and we do not strictly require that an experiment is performed on $\calI$, provided that a suitable independence condition (given in Assumption \ref{as:unconf}) is satisfied.
We also refer to $\calI$ interchangeably as the source sample, source data, etc.

For each unit $i \in \calI$, we observe $(Y_i,D_i,X_i,A^\calI_{i1}, \ldots, A^\calI_{in^\calI})$, where $Y_i \in \bbR$ is the outcome of interest, $D_i \in \{0,1\}$ is the treatment indicator, and $X_i \in \calX$ is the vector of covariates.
We assume that $\calX$ is a finite set with $d_x = |\calX|$.
Here, $A^\calI_{ij}$ denotes the $(i,j)$-th element of $n^\calI \times n^\calI$ adjacency matrix $\bm{A}^\calI$.
We assume that $\bm{A}^\calI$ is fixed during the experiment and treat it as a non-stochastic object.
In addition, $\bm{A}^\calI$ does not have self-loops and may or may not be directed; i.e., $A^\calI_{ij} \ne A^\calI_{ji}$ is allowed.
The \textit{degree} of $i$, the number of network connections of $i$, is denoted as $G_i = \sum_{j \in \calI} A^\calI_{ij}$.
When $\bm{A}^\calI$ is directed, $G_i$ is interpreted as the out-degree of $i$. 
Because $\bm{A}^\calI$ is treated as non-stochastic, so is the degree $G_i$.  
Let $\calG$ denote the finite set of possible degree values and write $d_g = |\calG|$.
For a generic $n^\calI$-dimensional treatment assignment $\bm{d}^\calI \in \{0,1\}^{n^\calI}$, define the number of treated peers for $i$ by $S_i(\bm{d}^\calI) \coloneqq \sum_{j \in \calI} A^\calI_{ij} d_j$ and denote its realized value by $S_i = S_i(\bm{D}^\calI)$, where $\bm D^\calI = \{D_i : i \in \calI\}$.

Next, we introduce the \textit{exposure mapping} $E_i = e(S_i,G_i)$, where $e: \calG \times \calG \to \calE$ is a deterministic function chosen by the researcher.
The exposure mapping summarizes peer-treatment impacts into lower dimensional statistics.
Since individuals usually have their own unique social networks, it is generally impossible to identify meaningful causal parameters without dimensionality reduction of the interference structure, and exposure mapping is the standard approach in the literature (e.g., \citealp{aronow2017estimating, aronow2021spillover, leung2024identifying}).
Typical choices of exposure mapping include $e(S,G)=S$, $e(S,G)=S/G$, and $e(S,G)=\bm{1}\{S>0\}$.
In this study, we assume that the exposure mapping is correctly specified (in the sense of Assumption \ref{as:tl}(i) below) as a function of $(S, G)$.\footnote{
    When the exposure mapping is misspecified, the resulting estimates generally exhibit biased causal interpretations; see \cite{leung2024identifying}.
    To mitigate the misspecification problem, \cite{hoshino2023randomization} propose a specification test for the exposure mapping.
    }
    
The individuals of primary interest are not those in the experimental sample but those in the target data $\calJ$, whose size is $n^\calJ \coloneqq |\calJ|$.  
We assume that either the same set of covariates $X_j \in \calX$ as in $\calI$ is observed for all $j \in \calJ$, or the distribution of $X$ over $\calJ$ is known (i.e., the proportion of each $x \in \calX$ is observed).  
The former is reasonable if $X$ consists of socioeconomic characteristics that policymakers can access through official surveys.  
\cite{miao2024transfer} consider transfer learning when only a subset of $X$ is observed for the target data.
The latter case corresponds to situations, for example, where $X_j$ contains private information or where $\calJ$ is so large that individual-level data cannot be collected and only their distributions are publicly available to researchers.
In both cases, the covariates for the target data are treated as given.
As mentioned in the introduction, the network $\bm A^\calJ$ in the target data is assumed to be unobserved.  
Although partial knowledge of $\bm A^\calJ$ can help tighten the bounds on policy effects, we focus on the case where $\bm A^\calJ$ is completely unknown for clarity of presentation.

\bigskip

Now, we introduce the following transferability assumption.
\begin{assumption}[Transferability]\label{as:tl}
\begin{itemize}
    \item[(i)] For both data $\calI$ and $\calJ$, the outcome $Y$ is generated as 
    \begin{align}\label{eq:model}
        Y = y(D, E, G, X, \epsilon),
    \end{align} 
    where $\epsilon$ is an unobserved disturbance term of arbitrary dimension.
    \item[(ii)] For all $i \in \calI$, $j \in \calJ$, and $(d,e,g,x) \in \{0,1\} \times \calE \times \calG \times \calX$,
    \begin{align}
        \mu(d,e,g,x) \coloneqq \bbE^\calI[Y_i(d,e) \mid G_i = g, X_i = x] = \bbE^\calJ[Y_j(d,e) \mid G_j = g, X_j = x],
    \end{align}
    where $Y(d,e) \coloneqq y(d,e,G,X,\epsilon)$ denotes the potential outcome when $(D,E) = (d,e)$.
\end{itemize}
\end{assumption}

Assumption \ref{as:tl} requires a certain degree of similarity between the source and target data to ensure the transferability of results.  
Specifically, condition (i) imposes two main restrictions.  
First, the exposure mapping $E = e(S,G)$ must be correctly specified.\footnote{
    Since $E$ is fully determined by $(S,G)$, we may rewrite \eqref{eq:model} as $Y = y(D,S,G,X,\epsilon)$.  
    A similar model specification can be found in \cite{leung2020treatment}.  
    However, except when $\calG$ is a very small set, a fully nonparametric regression on $(S,G)$ is unrealistic, so the use of an exposure mapping will eventually be required in practice.
    We express our model in the form of \eqref{eq:model} to highlight this point.
}
Note that a correct exposure mapping is generally not unique; any mapping consistent with \eqref{eq:model} can be employed.
However, for estimation efficiency, it is preferable to use a "coarser" mapping (in the sense of  \cite{hoshino2023randomization}).
Second, the outcome may depend on the network structure, but only through the degree $G$.
This is motivated by possible heterogeneity in treatment and spillover effects with respect to $G$.
For example, when $e(S,G)=S/G$ is used, we wish to distinguish between having exactly one friend, who is treated, and having many friends, all of whom are treated.
In addition to $G$, if desired, our approach allows to include other "node-level" network covariates (e.g., centrality, local clustering) in the model, as in \cite{lin2017estimation}; however, the resulting prediction bounds will be much larger relative to the present specification.
Also note that the model cannot incorporate "network-level" statistics as covariates; an extreme case is $Y = y_{\bm A}(D,E,G,X,\epsilon)$.
In such cases, it is impossible to generalize results from a single network to another network without additional structural assumptions.

Condition (ii) is our main transferability assumption and parallels the covariate-shift assumption in the transfer learning literature.  
It states that the relationship between the potential outcomes and the covariates $(G,X)$ is the same in the source and target data, although the distributions of these variables may differ.  
Given condition (i), condition (ii) holds if the conditional distributions of $\epsilon_i$ ($i \in \calI$) and $\epsilon_j$ ($j \in \calJ$) given $(G,X)$ are identical.  
This is plausible when $\calI$ and $\calJ$ are drawn from the same population; for example, $\calI$ is a village where an experiment was conducted, and $\calJ$ comprises all other villages in the same province, as in \cite{cai2015social}.
If we can additionally assume the additive separability: $y(D,E,G,X,\epsilon)=\mu(D,E,G,X)+\epsilon$, then condition (ii) reduces to requiring only that the error terms have mean zero conditional on $(G,X)$. 

\bigskip

It is important to note that merely estimating $\mu(d,e,g,x)$ may not necessarily be informative for evaluating the social impact of a specific policy (i.e., a treatment rule) among the target data.  
This is because in order to evaluate a given treatment rule, we need to determine not only each unit's own treatment status $d$, but also the exposure value $e$.  
However, the exposure $e$ is not identifiable in the absence of network information, in general.

With this in mind, we now introduce our main causal parameter of interest.  
Let $Y_i(\bm d^{\calI})$ denote the potential outcome when $\bm D^{\calI} = \bm d^{\calI}$.  
Observe that the two potential outcome notations are related in the following manner: $Y_i(\bm d^{\calI}) = Y_i(d_i, e(S_i(\bm d^{\calI}), G_i))$.
Then, the total treatment effect (TTE) for unit $i \in \calI$ is defined as
\begin{align}
  \tau_i
  &\coloneqq Y_i(\bm 1_{n^{\calI}}) - Y_i(\bm 0_{n^{\calI}}) \\
  &= Y_i(1, e(G_i, G_i)) - Y_i(0, e(0, G_i)).
\end{align}
The TTE for $j \in \calJ$ is similarly defined.  
The TTE is interpreted as the individual-level effect of a policy that assigns all units in the same network to treatment. 
There is a large literature on statistical inference for parameters related to the TTE (e.g., \citealp{chin2019regression, yu2022estimating, ugander2023randomized, faridani2024linear}).
In particular, \cite{yu2022estimating} is conceptually related to our study in that they also consider estimation under unknown networks.\footnote{
    Their method, like ours, does not require knowledge of the network structure.
    However, it assumes that the direct and interference effects are additively separable and that researchers have prior knowledge of the average baseline outcome.
    The approach of \cite{faridani2024linear} also allows for settings without precise information about network connections.
    However, they assume that there is a known distance measure, such that spillover effects decay in a power of this distance.
}
A notable fact is that the TTE depends on the degree of $i$ but not on the other network statistics.  

Because we can observe only one potential outcome for each individual, individual TTEs are not computable.
Hence, we adopt the average TTE conditioned on the degree and covariates, which we refer to as the ATTE, as our main parameter of interest:
\begin{align}
    \kappa^{\calJ}
    \coloneqq \frac{1}{n^\calJ} \sum_{j \in \calJ} \bbE^{\calJ}[\tau_j \mid G_j, X_j].
\end{align}
By Assumption \ref{as:tl}(ii),
\begin{align}
    \bbE^\calJ [\tau_j \mid G_j, X_j]
    &= \sum_{x \in \calX} \sum_{g \in \calG}
       \bbE^{\calJ}[\tau_j \mid G_j = g, X_j = x]
       \bm{1}\{G_j = g, X_j = x\} \\
    &= \sum_{x \in \calX} \sum_{g \in \calG}
       ( \mu(1, e(g,g), g, x)
       - \mu(0, e(0,g), g, x))
       p^{\calJ}(x) \frac{\bm{1}\{G_j = g, X_j = x\}}{p^{\calJ}(x)} ,
\end{align}
where $p^{\calJ}(x)$ is the proportion of units with covariate value $x$ in the target data, which is assumed to be known.
When we can observe $X_j$ for all $j \in \calJ$, we set $p^{\calJ}(x) = (n^\calJ)^{-1} \sum_{j \in \calJ} \bm{1}\{X_j = x\}$.
Moreover, letting $\pi^\calJ(g,x)$ be the conditional degree distribution given $X = x$:
\begin{align}
  \pi^{\calJ}(g, x)
  &\coloneqq \frac{1}{n^\calJ}
              \sum_{j \in \calJ}
              \frac{\bm{1}\{G_j = g, X_j = x\}}{p^{\calJ}(x)},
\end{align}
we can write the ATTE as
\begin{align}\label{eq:atte}
    \kappa^{\calJ}
    = \sum_{x \in \calX} \sum_{g \in \calG}(\mu(1, e(g,g), g, x) - \mu(0, e(0,g), g, x)) p^{\calJ}(x) \pi^{\calJ}(g, x).
\end{align}
As shown here, if $p^{\calJ}$ is known, we do not need to collect individual covariates to compute $\kappa^{\calJ}$.
However, since $G_j$'s are unobserved, $\pi^{\calJ}(g, x)$ is also unknown, so $\kappa^{\calJ}$ cannot be computed directly.
For this issue, the next section introduces a sensitivity analysis framework with respect to the uncertainty of $\pi^{\calJ}$.

\begin{remark}[Separating the direct and spillover effects]
It is easy to see that the ATTE can be decomposed into a direct effect and a spillover effect:
$\kappa^{\calJ}=\kappa^{\calJ}_{\text{direct}}+\kappa^{\calJ}_{\text{spill}}$, where
\begin{align}
  \kappa^{\calJ}_{\text{direct}}
  &= \sum_{x \in \calX} \sum_{g \in \calG}
     \bigl(\mu(1,e(g,g),g,x)-\mu(0,e(g,g),g,x)\bigr)p^{\calJ}(x)\pi^{\calJ}(g,x), \\
  \kappa^{\calJ}_{\text{spill}}
  &= \sum_{x \in \calX} \sum_{g \in \calG}
     \bigl(\mu(0,e(g,g),g,x)-\mu(0,e(0,g),g,x)\bigr)p^{\calJ}(x)\pi^{\calJ}(g,x).
\end{align}
Applying our proposed method, we can construct bounds for $\kappa^{\calJ}_{\text{direct}}$ and
$\kappa^{\calJ}_{\text{spill}}$ separately.  
However, caution is needed in interpreting these
quantities.
Note that $\sum_{x \in \calX} \sum_{g \in \calG}\mu(0,e(g,g),g,x)p^{\calJ}(x)\pi^{\calJ}(g,x)$ represents the average of conditional mean outcomes when all units are untreated but at the same time all of their peers are treated, which is a logical contradiction.
Therefore, $\kappa^{\calJ}_{\text{direct}}$ and $\kappa^{\calJ}_{\text{spill}}$ are not, by themselves, representing "policy effects" of any implementable policy.
\end{remark}

Lastly in this section, we discuss the identification of $\mu(d,e,g,x)$.  
The following assumption is plausible when the source data are obtained through an RCT.
\begin{assumption}[Unconfoundedness]\label{as:unconf}
$\epsilon_i \indep \bm{D}^{\calI} \mid G_i, X_i$ for all $i \in \calI$.
\end{assumption}
Assumption \ref{as:unconf} ensures that, conditional on $(G_i,X_i)$, the potential outcomes $\{Y_i(d,e)\}$ are independent of the realized $(D_i,E_i)$.
Since $\mu(d,e,g,x)$ is estimated using only the source data, this assumption is not required for the target data.  
Under Assumption \ref{as:unconf},
\begin{align}
    \bbE^{\calI}\left[ Y_i \mid D_i = d, E_i = e, G_i = g, X_i = x \right]
    & = \bbE^{\calI}\left[ Y_i(d,e) \mid D_i = d, E_i = e, G_i = g, X_i = x \right] \\
    &= \mu(d,e,g,x).
\end{align}
This implies that $\mu(d,e,g,x)$ is nonparametrically identifiable when the event $\{D_i = d, E_i = e, G_i = g, X_i = x\}$ occurs with positive probability.

\section{The Linear Programming Problem}\label{sec:LP}

\subsection{A linear-programming characterization of ATTE}
 
As shown in \eqref{eq:atte}, in order to compute the ATTE $\kappa^\calJ$ directly, we need the information of $\pi^\calJ(g,x)$, which is unavailable by assumption.  
Instead, suppose the researcher has a candidate baseline conditional degree distribution $\pi_x^* \in \calP(\calG)$, where $\calP(\calG)$ is the set of probability distributions whose support is a subset of $\calG$.  
There are several reasonable choices for the baseline distribution.  
The most natural option would be to use the degree distribution in the source data 
$\pi_x^*(g) = \pi^\calI(g,x) \coloneqq (n^{\calI})^{-1} \sum_{i \in \calI} \bm{1}\{G_i = g, X_i = x\}/p^\calI(x)$, where $p^\calI(x) \coloneqq (n^{\calI})^{-1} \sum_{i \in \calI} \bm{1}\{X_i = x\}$.
This choice is particularly advocated when the source and target data come from the same population.  
Another possibility is to learn a link-prediction model using any method with the source data $\{(X_i, A^\calI_{i1},\ldots,A^\calI_{in^{\calI}}): i \in \calI\}$, obtain a predicted adjacency matrix for $\calJ$, $\hat{\bm{A}}^\calJ$, and set $\pi_x^*$ to the conditional degree distribution on $\hat{\bm{A}}^\calJ$.
If the researcher has background knowledge about the target data from previous studies and observations, $\pi_x^*$ may instead be specified a priori.  

To quantify the distance between distribution functions in $\calP(\calG)$, this paper uses the Wasserstein distance.
\begin{definition}[$q$-Wasserstein distance]
The $q$-Wasserstein distance between $\pi \in \calP(\calG)$ and $\pi^* \in \calP(\calG)$ is given as follows ($q \in [1, \infty)$):
    \begin{align}
      \mathcal{W}_q(\pi, \pi^*)
      \coloneqq
      \left( \min_{\Gamma \in \Pi(\pi, \pi^*)}
      \sum_{(u,v)\in \calG^2}
        \Gamma(u,v) \bigl|u-v\bigr|^q\right)^{1/q},
    \end{align}
    where $\Pi(\pi, \pi^*)$ consists of all nonnegative matrices $\Gamma(u,v)$ satisfying
    \begin{align}
      \sum_{v \in \calG}\Gamma(u,v) = \pi^*(u),
      \quad
      \sum_{u \in \calG}\Gamma(u,v) = \pi(v).
    \end{align}
\end{definition}

The Kullback-Leibler divergence is a popular choice in sensitivity analysis for quantifying the distance to a reference distribution (e.g., \citealp{duchi2021learning, spini2021robustness, christensen2023counterfactual}).  
However, its greatest limitation lies in the requirement of absolute continuity, which significantly restricts the choice of reference degree distribution $\pi_x^*$.  
For example, if we set $\pi_x^*(g) = \pi^\calI(g,x)$, then, because $\calI$ is typically smaller than $\calJ$, the support of $\pi^\calI(g,x)$ is likely to be strictly contained in that of $\pi^\calJ(g,x)$.  
Consequently, $\pi^\calJ$ is not absolutely continuous with respect to $\pi^\calI$ and is therefore excluded from the candidate set of distributions.\footnote{
    Note that if the support of $\pi^\calI(g,x)$ is a strict subset of that of $\pi^\calJ(g,x)$, then it is impossible to nonparametrically estimate $\mu(d,e,g,x)$ on those $(g,x)$ values.
    In such cases, one eventually needs to perform inter- or extrapolation of the estimates by assuming a functional form such as in \eqref{eq:vc}.
}
In contrast, the Wasserstein distance can be computed for essentially any pair of distributions.  
Moreover, using the Wasserstein distance allows us to characterize the bounds on $\kappa^\calJ$ through a set of linear programming problems.  
For a more detailed discussion of the advantages of the Wasserstein distance over the Kullback--Leibler divergence in the context of distributionally robust optimization, see \cite{gao2023distributionally}.  

Next, we define the $(\delta, q)$-Wasserstein ball centered at $\pi^*$: 
\begin{align}
    \bbB(\pi^*, \delta, q) \coloneqq \{\pi \in \calP(\calG): \mathcal{W}_q(\pi, \pi^*) \le \delta\},
\end{align}
for a radius $\delta \in (0,\infty)$.
Then, the lower and the upper bounds for $\kappa^\calJ$ at a given Wasserstein radius $\delta$ can be formulated as follows, respectively: 
\begin{align}\begin{split}\label{eq:bounds}
    \underline{\kappa}_{\delta, q} & \coloneqq \sum_{x \in \calX} \left[ \min_{\pi_x \in \bbB(\pi_x^*, \delta, q)} \sum_{g \in \calG} m(g, x) \pi_x(g) \right] \\
    \overline{\kappa}_{\delta, q} & \coloneqq \sum_{x \in \calX} \left[ \max_{\pi_x \in \bbB(\pi_x^*, \delta, q)} \sum_{g \in \calG} m(g, x) \pi_x(g) \right] 
\end{split}\end{align}
where 
\begin{align}
    m(g, x) \coloneqq \left( \mu(1, e(g,g), g, x) - \mu(0, e(0,g), g, x) \right) p^\calJ(x).
\end{align}
Clearly, if $\pi^\calJ(\cdot, x) \in \bbB(\pi_x^*, \delta, q)$ for every $x \in \calX$, $\kappa^\calJ \in [\underline{\kappa}_{\delta, q}, \overline{\kappa}_{\delta, q}]$ holds.
In addition, $\pi^\calJ(\cdot, x) \in \bbB(\pi_x^*, \delta, q)$ holds for any baseline $\pi^*_x$ if we take sufficiently large $\delta$.

\begin{example}\label{exp:simple}
    To illustrate the bounds \eqref{eq:bounds}, we provide a toy example here.
    Suppose that there are no covariates and there are only two support points for the degree distribution: $\calG = \{0,1\}$.
    $m$ depends only on $g$, and we assume $m(0) \le m(1)$.
    For the baseline degree distribution, we set $\pi^*(g) = (\alpha^*)^g (1 - \alpha^*)^{1 - g}$.
    Then, for any Bernoulli distribution $\pi(g) = (\alpha)^g (1 - \alpha)^{1-g}$, setting $q = 1$, $\mathcal{W}_1(\pi, \pi^*) = |\alpha - \alpha^*|$ holds.
    Under this setup, the lower and the upper bounds can be obtained as follows:
    \begin{align}
        \underline{\kappa}_{\delta, 1}
        & = \min_{\alpha \in [0,1] \; : |\alpha - \alpha^*| \le \delta} (1 - \alpha) m(0) + \alpha m(1)  \\
        & = m(0) + \max\{0, \alpha^* - \delta\} (m(1) - m(0) ) \\
        \overline{\kappa}_{\delta, 1}
        & = \max_{\alpha \in [0,1] \; : |\alpha - \alpha^*| \le \delta} (1 - \alpha) m(0) + \alpha m(1) \\
        & = m(0) + \min\{1, \alpha^* + \delta\} (m(1) - m(0) ).
    \end{align}
    Hence, if the chosen $\delta$ is large enough, we will have the trivial bounds  $\underline{\kappa}_{\delta, 1} = m(0)$ and $\overline{\kappa}_{\delta, 1} = m(1)$. 
   
    Figure \ref{fig:example} presents the areas of $[\underline{\kappa}_{\delta, 1}, \overline{\kappa}_{\delta, 1}]$ when $m(0) = 0$, $m(1) = 1$, and $\pi^{\mathcal J}(g) = (0.4)^g (0.6)^{1 - g}$.
    The dotted horizontal line corresponds to the target parameter $\kappa^\calJ = 0.4$.
    It is evident from the left panel that, when the Wasserstein ball is centered at the true $\pi^{\mathcal J}(g)$, the interval $[\underline{\kappa}_{\delta,1}, \overline{\kappa}_{\delta,1}]$ contains $\kappa^{\mathcal J}$ for any value of $\delta > 0$.
    As the middle and right panels illustrate, even when the reference probability distribution $\pi^*$ deviates from the true $\pi^\calJ$, increasing $\delta$ sufficiently large still ensures the coverage of $\kappa^\calJ$.
    
    \begin{figure}[ht]
      \centering              
      \includegraphics[width=16cm]{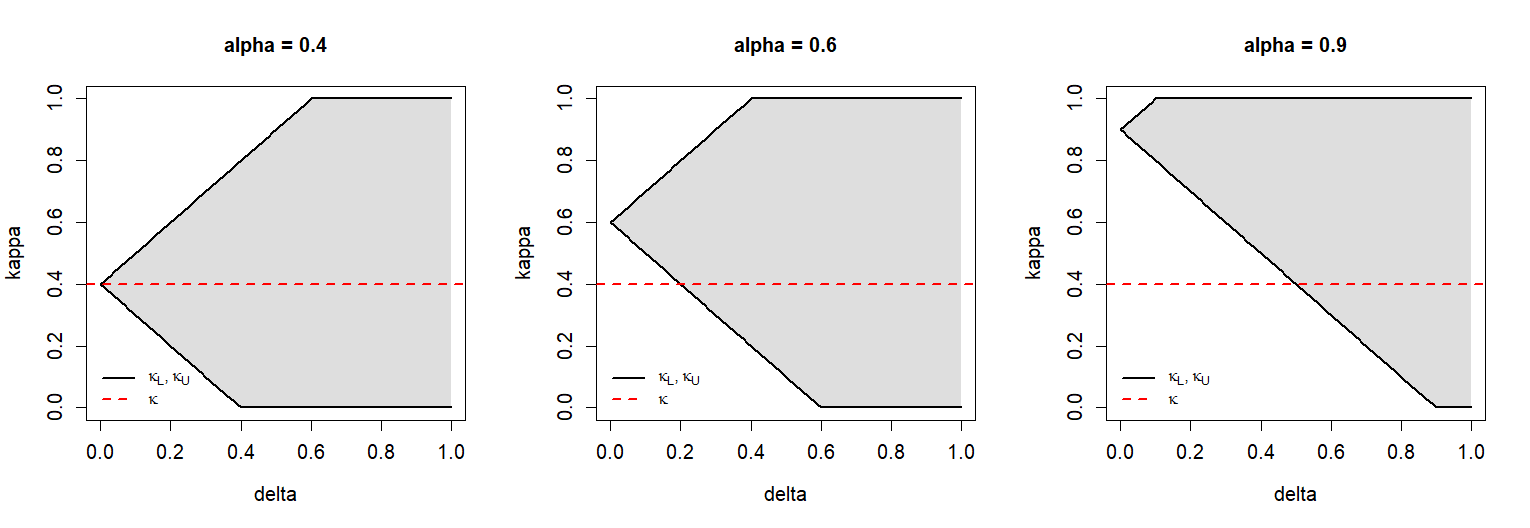}
    
      \caption{Upper and lower bounds of $\kappa^\mathcal{J}$}
      \label{fig:example}
      \bigskip  
      \footnotesize\flushleft
      Notes: $\pi^\calJ=\text{Bernoulli}(0.4)$, $m(1)=1$, and $m(0)=0$.
      (Left) $\alpha^*=0.4$.
      (Middle) $\alpha^*=0.6$.
      (Right) $\alpha^*=0.9$.
    \end{figure}
\end{example}
  
Hereinafter, since every minimization problem can be converted into a maximization problem by changing the sign of the objective function, we mainly focus on the computation of the upper bound $\overline{\kappa}_{\delta, q}$.
For completeness, the estimation and inference for the lower bound $\underline{\kappa}_{\delta, q}$ are summarized in Appendix \ref{app:lower}.

Our goal is to maximize the following objective function: $\sum_{x \in \calX} \sum_{g \in \calG} m(g, x)\pi_x(g)$ with respect to $\pi_x$ subject to $\pi_x \in \bbB(\pi_x^*, \delta,q)$ for each $x \in \calX$.
Note that since $\sum_{u \in \calG} \Gamma_x(u,v) = \pi_x(v)$, we may write $\sum_{g \in \calG} m(g, x)\pi_x(g)  = \sum_{u,v \in \calG^2} \Gamma_x(u,v) m(v, x)$.
In addition, restricting the parameter space to the Wasserstein ball $\bbB(\pi_x^*, \delta,q)$ is equivalent to satisfying the following set of linear equalities and inequalities: $\sum_{v \in \calG}\Gamma_x(u, v) = \pi_x^*(u)$, $\sum_{u, v \in \calG^2} \Gamma_x(u, v) \bigl|u - v\bigr|^q \le \delta^q$, and $\Gamma_x(u, v) \ge 0$.
Consequently, the upper bound $\overline{\kappa}_{\delta, q}$ corresponds the objective value of the following linear program:
\begin{align}\label{eq:lp}
\begin{split}
    &\text{maximize } \sum_{x \in \calX} \sum_{u,v \in \calG^2} \Gamma_x(u,v) m(v, x) \\
    &\text{subject to } \sum_{v \in \calG}\Gamma_x(u, v) = \pi_x^*(u), \sum_{u, v \in \calG^2} \Gamma_x(u, v) \bigl|u - v\bigr|^q \le \delta^q, \Gamma_x(u, v) \ge 0, \forall\, (x, u, v) \in \calX \times \calG^2 
\end{split}
\end{align}
    
\begin{remark}[Non-uniqueness of the solution]\label{rem:nonuniq}
    Since the parameter space for $\Gamma_x$ is a compact convex subset of the probability simplex, the optimal value in the problem \eqref{eq:lp} exists uniquely.
    However, as in typical linear programming problems, the solution that attains the optimal value is not unique in general.\footnote{
        For example, consider the following setup without covariates: $\calG = \{1,2,3\}$, $(m(1),m(2),m(3)) = (1,2,3)$, $(\pi^*(1), \pi^*(2), \pi^*(3)) = (1/2, 1/2, 0)$, $\delta = 1$, and $q = 1$.
        Then, the optimal $\pi$ is given for example by $(\pi(1), \pi(2), \pi(3)) = (0, 1/2, 1/2)$, which can be achieved by two different transference plans: 
        $\Gamma^{(1)} =
        \begin{bmatrix}
        0 & 0 & \tfrac{1}{2} \\
        0 & \tfrac{1}{2} & 0 \\
        0 & 0 & 0
        \end{bmatrix}$ and
        $\Gamma^{(2)} =
        \begin{bmatrix}
        0 & \tfrac{1}{2} & 0 \\
        0 & 0 & \tfrac{1}{2} \\
        0 & 0 & 0
        \end{bmatrix}$.
    }
    Note that once the target distribution $\pi_x$ is fixed for each $x \in \calX$, the optimal transference plan can be found uniquely when $q > 1$ (see, e.g., Theorem 1.5.1 in \cite{panaretos2020invitation}).
    For the uniqueness of $\pi_x$, since the objective function is linear, the solution $\pi_x$ will be unique if the Wasserstein ball $\bbB(\pi_x^*, \delta, q)$ were strictly convex, which is not true in general in our setting.
\end{remark}

Despite the non-uniqueness of the solution to problem \eqref{eq:lp}, one might still wish to exemplify specific network structures that attain the maximum or minimum objective value. 
Note, however, that the degree distribution obtained from \eqref{eq:lp} need not be \textit{graphic}; that is, it is not always possible to realize an arbitrary degree distribution with a simple graph.
In graph theory, the Erd\"{o}s--Gallai theorem provides a simple necessary and sufficient condition for a sequence of positive integers to be graphic (see, e.g., \citealp{tripathi2010short}).
When this condition is met, one can generate such graphs using some computational algorithms.\footnote{
    For example, the \texttt{igraph} \texttt{R} package offers the function \texttt{realize\_degseq} that can be used for this purpose.
    }
Meanwhile, even when the obtained degree distribution is not graphic, it is still possible to construct a graph whose expected degree sequence matches the given degree distribution, for instance, by employing the Chung-Lu model (see, e.g., 4.1.5 of \citealp{jackson2008social}).

\begin{remark}[Interpretation of $\delta$]
Interpreting the Wasserstein neighbourhood size $\delta$ in practice is a central issue in sensitivity analysis.
One possible approach is to split the source data into disjoint networks.  
For example, in a school experiment, students' friendship networks are often disjoint across grades.
Then, by computing the Wasserstein distance between the degree distribution of one grade and that of another, we obtain a typical discrepancy value $\hat\delta$ between degree distributions drawn from the same population.  
If the source and target data are believed to come from a similar population, we may then set $\delta$, conservatively, for example $\delta\in(0,2\hat\delta \, ]$.  

\end{remark}

The number of variables in the linear program \eqref{eq:lp} is $d_x d_g^2$.  
Although the problem can be simplified by decomposing it into $d_x$ sub-linear programs, some computational effort may still be required when $d_g$ is large.
Fortunately, the dual problem of \eqref{eq:lp} can be easily derived.

\begin{proposition}[Dual problem]\label{prop:dual}
Suppose that $m(v,x)$ is bounded uniformly in $(v,x) \in \calG \times \calX$.
Then, for any $q \in [1, \infty)$ and $\delta > 0$,
\begin{align}\label{eq:dual}
\begin{split}
    \overline{\kappa}_{\delta, q} 
    & = \sum_{x \in \calX} \left[ \min_{\lambda_x \ge 0}\left\{ \lambda_x \delta^q + \sum_{u \in \calG} \max_{v \in \calG}\{m(v, x) - \lambda_x |u-v|^q\} \pi^*_x(u)  \right\} \right].
\end{split}
\end{align}
\end{proposition}

Proposition \ref{prop:dual} shows that the optimal value of the primal linear program \eqref{eq:lp} can be obtained by solving $d_x$ separate univariate minimization problems.
The derivation of \eqref{eq:dual} is provided in Appendix \ref{app:dual}.
For a formal proof in a more general setting, see Theorem 1 of \cite{blanchet2019quantifying} or Theorem 1 of \cite{gao2023distributionally}. 
As an illustration, the dual problem for Example \ref{exp:simple} is presented in Appendix \ref{app:exp}.

\subsection{Examples of degree distributions}\label{subsec:degree}

When the researcher has prior knowledge about the network structure in the target data, the baseline $\pi_x^*$ can be chosen based on it.
For example, when links are believed to exist independently with each other with equal probability (i.e., an Erd\H{o}s--R\'enyi graph), the degree distribution of a large network can be approximated by a Poisson distribution.
However, many empirical networks are known to deviate substantially from the Poisson distribution (e.g., \citealp{albert2002statistical}).
For example, across a wide range of scientific areas, a power-law distribution (i.e., $\pi(g) \sim g^{-c}$ for some $c > 0$) often serves as a good approximation of the observed degree distribution (e.g., \citealp{kolaczyk2009statistical}).

\begin{figure}[ht]
  \centering
  \begin{minipage}[t]{0.48\textwidth}
    \centering
    \includegraphics[width=\textwidth]{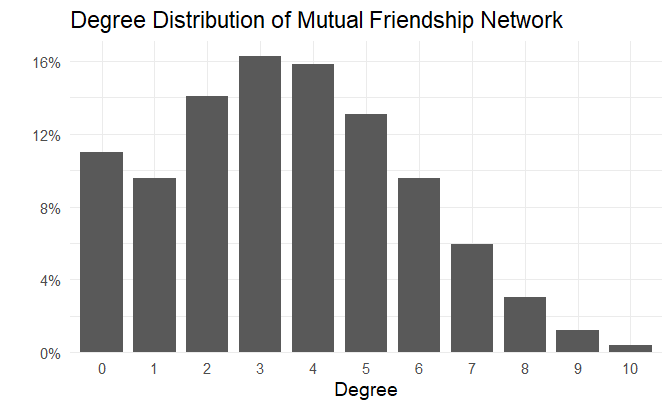}
  \end{minipage}
  \hfill
  \begin{minipage}[t]{0.48\textwidth}
    \centering
    \includegraphics[width=\textwidth]{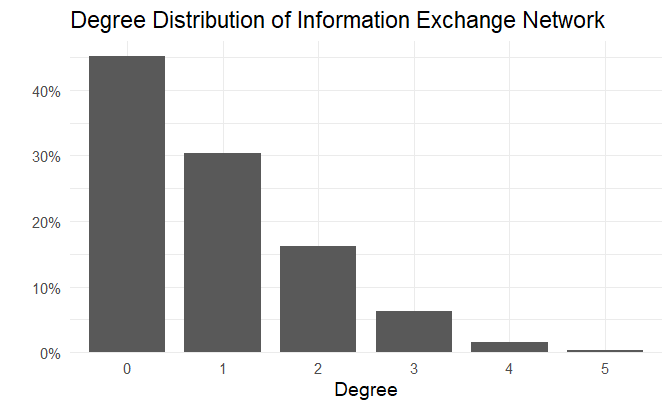}
  \end{minipage}

  \caption{Real data examples of degree distribution}
  \label{fig:deg_real}

  \bigskip

  \footnotesize\flushleft
  Notes: (Left) Mutual friendship network among students: data source \cite{paluck2016changing}.
  (Right) Mutual information exchange network among farmers: data source \cite{cai2015social}.
\end{figure}

Meanwhile, in social relationship networks, extremely large degrees are rarely observed in practice.  
Figure \ref{fig:deg_real} shows the degree distributions of a mutual friendship network among students and a bilateral information-exchange network among farmers, created from \cite{paluck2016changing} and \cite{cai2015social}, respectively.  
In both cases, we assume that there is a link only when the two individuals nominate each other as partners.  
As indicated in the left panel, the number of closest school friends peaks at about three or four.  
In the information exchange network among farmers, a large share of farmers has no such partner.

These observations suggest that, depending on the type of data, its degree distribution may follow a typical shape pattern such as unimodality, monotonicity, or symmetry.
Explicitly imposing the shape restrictions on the candidate degree distributions can yield tighter prediction bounds.
For example, in the case of monotonicity as in the right panel of Figure \ref{fig:deg_real}, we can add the linear inequality constraints $\pi_x(g_1) \ge \pi_x(g_2)$ for all $g_1 < g_2$ directly into the linear program \eqref{eq:lp}.

\section{Estimation and Asymptotic Properties}\label{sec:estimation}

\subsection{Estimation}\label{subsec:vcoef}

The linear program in \eqref{eq:lp} is not feasible because $m(g,x)=(\mu(1,e(g,g),g,x)-\mu(0,e(0,g),g,x))p^\calJ(x)$ is unknown.  
Nonparametrically estimating $\mu(d,e,g,x)$ is unrealistic due to the curse of dimensionality, except when the sample size $n^\calI$ is extremely large.  
Therefore, we would need to introduce additional functional-form restrictions on the outcome equation $y(d,e,g,x,\epsilon)$ in most applications.  
Although many specifications could be considered, we adopt the following varying-coefficient model as a typical candidate:
\begin{align}\label{eq:vc}
  y(d,e,g,x,\epsilon) = w(d,e,g)^\top \beta(x) + \epsilon,
\end{align}
where $w: \{0,1\} \times \calE \times \calG \to \bbR^{d_w}$ is a pre-specified basis function, and $\epsilon$ is a scalar error term.  
Then, under this specification, we only need to estimate the coefficient functions $\beta(x)$ to recover $m(g,x)$.

For the estimation of $\beta(x)$, we adopt the kernel weighted regression approach proposed by \cite{li2010smooth}.
Recalling that the covariates $X$ are discrete variables,
we partition $X$ into $d_c$-dimensional categorical variables $X^c$ and $d_o$-dimensional ordered variables $X^o$ ($d_c + d_o = d_x$).
Then, define the kernel weight function for discrete covariates as follows: $L_{i,b}(x) \coloneqq \prod_{j = 1}^{d_c} L^c_{ji,b}(x^c) \prod_{k = 1}^{d_o} L^{o}_{ki,b}(x^o)$, where
\begin{align}       
    L^{c}_{ji,b}(x^c) & \coloneqq \bm{1}\{X^c_{ji} = x_j^c\} + \bm{1}\{X^c_{ji} \neq x_j^c\} b_c \\
    L^{o}_{ki,b}(x^o) & \coloneqq \bm{1}\{X^o_{ki} = x_k^o\} + \bm{1}\{X^o_{ki} \neq x_k^o\} b_o^{|X^o_{ki} - x_k^o|},
\end{align}
$x = (x^c, x^o)$, $x^c = (x_1^c, \ldots, x_{d_c}^c)$, $x^o = (x_1^o, \ldots, x_{d_o}^o)$, with bandwidths $b = (b_c, b_o) \equiv (b_{c, n^\calI}, b_{o, n^\calI}) \in [0,1]^2$.
Our estimator of $\beta(x)$ is given by
\begin{align}\label{eq:coef}
    \hat \beta(x) \coloneqq \left( \frac{1}{n^\calI} \sum_{i \in \calI} W_i W_i^\top L_{i,b}(x) \right)^{-1} \frac{1}{n^\calI} \sum_{i \in \calI} W_i Y_i L_{i,b}(x),
\end{align}
where $W_i = w(D_i, E_i, G_i)$.
Then, $m(g, x)$ can be estimated by $\hat m(g, x) \coloneqq  z(g, x)^\top \hat \beta(x)$, where 
\begin{align}
    z(g, x) \coloneqq p^\calJ(x) \{ w(1, e(g,g), g) - w(0, e(0,g), g) \}.
\end{align}
Finally, by replacing $m$ in \eqref{eq:lp} by $\hat m$, we can estimate $\overline{\kappa}_{\delta, q}$ by
\begin{align}\begin{split}
    & \hat{\overline{\kappa}}_{\delta, q} \coloneqq \sum_{x \in \calX} \left[ \max_{\Gamma_x} \sum_{u,v \in \calG^2} \Gamma_x(u,v) \hat m(v, x) \right] \\
    &\text{subject to } \sum_{v \in \calG}\Gamma_x(u, v) = \pi_x^*(u), \sum_{u, v \in \calG^2} \Gamma_x(u, v) \bigl|u - v\bigr|^q \le \delta^q, \Gamma_x(u, v) \ge 0, \forall\, (x, u, v) \in \calX \times \calG^2  
\end{split}\end{align}
Of course, one may alternatively solve the dual problem \eqref{eq:dual} by putting $\hat m$ in the place of $m$.
We can similarly obtain $\hat{\underline{\kappa}}_{\delta, q}$, whose definition should be clear from the context.

\subsection{Asymptotic properties}

In this subsection, we derive the asymptotic distribution of $\hat{\overline{\kappa}}_{\delta,q}$ and present a wild bootstrap procedure for approximating the distribution.  
We begin by stating the asymptotic distributions of $\hat\beta$ and $\hat m$ in the next proposition.  
Since these results are not quite new and depend heavily on the model specification in \eqref{eq:vc}, all assumptions and detailed discussion are relegated to Appendix \ref{app:m}.  
The definitions of the asymptotic covariance matrices are also provided there.

\begin{proposition}[Asymptotic normality of $\hat \beta$ and $\hat m$]\label{prop:m}
    Suppose that Assumption \ref{as:vc} in Appendix \ref{app:m} holds.
    Then, 
    \begin{align}
    \text{(i)} \quad 
    & \sqrt{n^\calI} \left( \hat \beta(x) - \beta(x) \right)  \overset{d}{\to} N\left( \bm{0}_{d_w}, (\Sigma_\calI(x))^{-1} \Omega_\calI(x) (\Sigma_\calI(x))^{-1} \right) \text{ for each $x \in \calX$},\\
    \text{(ii)} \quad 
    & \sqrt{n^\calI} (\hat{\bm m} - \bm{m}) \overset{d}{\to} N\left( \bm{0}_{d_x d_g}, \bm{Z} \bm{J}_\calI \bm{\Omega}_\calI \bm{J}_\calI \bm{Z}^\top \right),
    \end{align}
    where $\bm m = (m(v_1, x_1), \ldots, m(v_{d_g}, x_1), \ldots, m(v_1, x_{d_x}), \ldots, m(v_{d_g}, x_{d_x}))^\top$, and $\hat{\bm m}$ is defined similarly.
\end{proposition}

We now turn to the asymptotic distribution of $\hat{\overline{\kappa}}_{\delta, q}$.
By the fundamental theorem of linear programming, an optimal $\Gamma_x$ for each $x \in \calX$ can be found among the set of basic feasible solutions of \eqref{eq:lp}; that is, the "corners" of the feasible set for $\Gamma_x$ satisfying all equality and inequality constraints in \eqref{eq:lp}.
We denote this set by $\calB_{\delta, q, x}$.
Let $\calS^*_{\delta, q, x}$ denote the set of maximizers:
\begin{align}
\calS^*_{\delta, q, x}
&\coloneqq \argmax_{\Gamma \in \calB_{\delta, q, x}} \sum_{u, v \in \calG^2} \Gamma(u, v) m(v, x).
\end{align}
Furthermore, define $\bbG = (\bbG(v_1, x_1), \ldots, \bbG(v_{d_g}, x_1), \ldots, \bbG(v_1, x_{d_x}), \ldots, \bbG(v_{d_g}, x_{d_x}))$ as a $d_x d_g$-dimensional multivariate normal random variable with mean zero and covariance matrix $\bm{Z} \bm{J}\calI \bm{\Omega}\calI \bm{J}_\calI \bm{Z}^\top$.

\begin{theorem}[Asymptotic distribution of $\hat{\overline{\kappa}}_{\delta, q}$]\label{thm:dist}
Suppose that Assumption \ref{as:vc} in Appendix \ref{app:m} holds.
Then, 
\begin{align}
    \sqrt{n^\calI}\left( \hat{\bar \kappa}_{\delta, q} - \bar \kappa_{\delta, q}\right) 
    & \overset{d}{\to} \sum_{x \in \calX} \left[ \max_{\Gamma_x \in \calS_{\delta, q, x}^*} \sum_{u,v \in \calG^2} \Gamma_x(u,v) \bbG(v, x) \right].
\end{align}
\end{theorem}
Theorem \ref{thm:dist} states that the limiting distribution of the upper-bound estimator is not pivotal, but can be numerically simulated through $\bbG$ to estimate the asymptotic critical values.
A similar result to our theorem can be found in \cite{bhattacharya2009inferring}.

To estimate the critical value at a given significance level, a natural approach would proceed as follows.  
First, we estimate $\calS^*_{\delta, q, x}$ by 
\begin{align}\label{eq:Shat}
    \hat{\calS}^*_{\delta, q, x} \coloneqq 
    \left\{ \Gamma \in \calB_{\delta, q, x} : 
    \sum_{u,v \in \calG^2} \Gamma(u,v) \hat{m}(v, x) 
    \ge \hat{\bar{\kappa}}_{\delta, q, x} - a \right\},
\end{align}
for some threshold parameter $a \equiv a_{n^\calI}$ tending to zero, where $\hat{\bar{\kappa}}_{\delta, q, x} \coloneqq \max_{\Gamma \in \calB_{\delta, q, x}} \sum_{u, v \in \calG^2} \Gamma(u, v) \hat m(v, x)$.
Second, generate independent draws 
$\bbG^{(r)} \sim N\left( \bm{0}_{d_x d_g}, \bm{Z} \bm{J}_{\calI} \bm{\Omega}_{\calI} \bm{J}_{\calI} \bm{Z}^\top \right)$ 
for $r = 1, \ldots, R$, with sufficiently large $R$.
For each draw, compute $\xi_{\delta, q}^{(r)} \coloneqq \sum_{x \in \calX} \left[ \max_{\Gamma_x \in \hat{\calS}^*_{\delta, q, x}} \sum_{u,v \in \calG^2} \Gamma_x(u,v) \bbG^{(r)}(v, x) \right]$.
Finally, the $\alpha$-level critical value is estimated by the $(1 - \alpha)$ empirical quantile of $\{\xi_{\delta, q}^{(r)} : r = 1, \ldots, R\}$.

This approach is straightforward, and a similar method has been considered in \cite{bhattacharya2009inferring}.
However, note that to implement the second step above, we must consistently estimate the covariance matrix $\bm{Z} \bm{J}_{\calI} \bm{\Omega}_{\calI} \bm{J}_{\calI} \bm{Z}^\top$, which typically requires a heteroscedasticity and autocorrelation consistent (HAC) estimator.
In general, the accuracy of the normal approximation with a HAC-estimated covariance matrix is limited when the sample size is not large.

Alternatively to the normal approximation with a HAC-estimated covariance, following \cite{fang2019inference}, this paper considers a bootstrap procedure.
In particular, since the data may exhibit cross-sectional dependence, we adopt the wild bootstrap approach by \cite{conley2023bootstrap}.
Specifically, to capture the dependence among units, we consider a setup similar to  \cite{kelejian2007hac}, \cite{kim2011spatial}, and \cite{conley2023bootstrap}.
That is, we assume that there is a socio-economic distance measure $\Delta_{ij}$ such that the dependence between $i$ and $j$ becomes stronger as $\Delta_{ij}$ becomes smaller.\footnote{
    If it is believed that the dependence is only through network link connections, we can alternatively use \cite{kojevnikov2021bootstrap}'s network wild bootstrap approach.
    The socio-economic distance-based approach considered here has the advantage of flexibility in the choice of distance measure, so that we can allow dependence of individuals even when they are apart in the network.
}
Although $\Delta_{ij}$ may be unobservable, an approximation $\tilde \Delta_{ij} = \Delta_{ij} + \nu_{ij}$ is available, where $\nu_{ij}$ is a measurement error.
Let $K: \bbR \to [-1, 1]$ be a real-valued kernel function, and define the matrix $\bbK_\calI \coloneqq (K(\tilde \Delta_{ij}/d))_{i,j \in \calI}$, where $d \equiv d_{n^\calI}$ is a bandwidth parameter.
Further, assuming that $\bbK_\calI$ is positive semidefinite,\footnote{\label{foot:psd}
    The positive semidefinite-ness of $\bbK_\calI$ is not always guaranteed and heavily depends on the choice of the kernel function $K$.
    For more detailed discussion on this issue, see, for example, \cite{kelejian2007hac} and \cite{conley2023bootstrap}.
} 
obtain its eigen-decomposition $\bbK_\calI = \Phi_\calI \Lambda_\calI \Phi_\calI^\top$, where $\Lambda_\calI$ is a diagonal matrix of the nonnegative eigenvalues of $\bbK_\calI$, and the columns of $\Phi_\calI$ are the corresponding orthonormal eigenvectors.
Now, we are ready to present our bootstrap procedure.

\begin{algorithm}[H]
\caption{Wild bootstrap procedure for inference on $\overline{\kappa}_{\delta,q}$}\label{algo:boot}
\begin{algorithmic}[1]
\State Estimate $\hat \beta(x)$ for all $x \in \calX$ using \eqref{eq:coef}
\State Compute the residual $\hat \epsilon_i \coloneqq Y_i - W_i^\top \hat \beta(X_i)$ for all $i \in \calI$
\For{$b = 1$ to $B$}
    \State Draw $\bm{\eta}^{(b)} = (\eta_1^{(b)}, \ldots, \eta_{n^{\calI}}^{(b)}) \sim \Phi_\calI \Lambda_\calI^{1/2} N(\bm{0}_{n^\calI}, I_{n^\calI})$
    \State Generate a bootstrap sample $\{(W_i, Y_i^{*(b)}): i \in \calI\}$, where $Y_i^{*(b)} \coloneqq W_i^\top \hat \beta(X_i) + \eta_i^{(b)} \hat \epsilon_i$
    \State Obtain $\hat \beta^{*(b)}(x)$ by the kernel weighted regression of $Y_i^{*(b)}$ on $W_i$ for all $x \in \calX$
    \State Compute $\hat{\bar \kappa}_{\delta,q}^{*(b)} \coloneqq \sqrt{n^{\calI}} \sum_{x \in \calX}  \left[ \max_{\Gamma_x \in \hat{\calS}_{\delta, q, x}^*} \sum_{u, v \in \calG^2} \Gamma_x(u,v) z(v, x)^\top (\hat \beta^{*(b)}(x) - \hat \beta(x)) \right]$
\EndFor
\State Compute the empirical $\alpha$ quantile $\hat \chi_{B,\alpha}$ of $\left\{\sqrt{n^\calI}(\hat{\bar \kappa}_{\delta,q}^{*(b)} - \hat{\bar \kappa}_{\delta, q}) : b = 1, \ldots, B\right\}$
\end{algorithmic}
\end{algorithm}

The validity of this bootstrap procedure is stated in the next proposition.
Again, the assumptions used here are all relegated to Appendix \ref{app:m}.

\begin{theorem}[Validity of the wild bootstrap]\label{thm:boot}
Suppose that Assumptions \ref{as:vc} and \ref{as:boot} in Appendix \ref{app:m} hold.
Then,
\begin{align}
    {\Pr}^*\left(\sqrt{n^\calI}(\hat{\bar \kappa}_{\delta,q}^* - \hat{\bar \kappa}_{\delta, q} ) \le s\right) = \Pr\left(\sqrt{n^\calI}( \hat{\bar \kappa}_{\delta, q} - \bar \kappa_{\delta, q}) \le s\right) + o_P(1)
\end{align}
uniformly in $s \in \bbR$, where ${\Pr}^*$ denotes the conditional probability given the source data.
\end{theorem}

Theorem \ref{thm:boot} implies that $\hat \chi_{B,\alpha}$ is a consistent estimator for the $\alpha$ quantile of $\sqrt{n^\calI}( \hat{\bar \kappa}_{\delta, q} - \bar \kappa_{\delta, q})$ as $B \to \infty$.
Therefore, the asymptotic $100(1 - \alpha)$\% confidence interval (CI) for $\bar \kappa_{\delta, q}$ can be obtained by
\begin{align}
    \calC_{1 - \alpha}(\bar \kappa_{\delta, q})
    \coloneqq \left[\hat{\bar \kappa}_{\delta, q} - \frac{\hat \chi_{B, 1 - \alpha/2}}{\sqrt{n^\calI}}, \:  \hat{\bar \kappa}_{\delta, q} - \frac{\hat \chi_{B, \alpha/2}}{\sqrt{n^\calI}} \right] .
\end{align}

\section{Monte Carlo Simulation}\label{sec:MC}

In this section, we examine the finite sample performance of the proposed wild bootstrap procedure through Monte Carlo simulations.
We consider the following data generating process:
\begin{align}
Y_i = \sum_{\ell=1}^6 W_{i\ell}\, b_\ell(X_{1i}, X_{2i}) + \epsilon_i, \;\; i \in \calI,
\end{align}
where $(W_{i1}, \ldots, W_{i6})^\top = w(D_i, e(S_i,G_i), G_i)$, $w(d,e,g) = (1, d, e, de, \log(g+1), e\log(g+1))$, $e(s,g) = s/g$, and
\begin{align}
\begin{array}{ll}
b_1(x_1,x_2) = 1, & b_2(x_1,x_2) = 1 + 0.5\Phi(x_1 + x_2),\\
b_3(x_1,x_2) = \Phi(x_1) + \Phi(x_2), & b_4(x_1,x_2) = \Phi(x_1) + \Phi(x_2),\\
b_5(x_1,x_2) = 0.5\exp\{-0.5(x_1 + x_2)\}, & b_6(x_1,x_2) = 0.5\exp\{-0.5(x_1 + x_2)\}.
\end{array}
\end{align}
The sample size is either $n^\calI = 400$ or $1200$.
The treatment variable and covariates are generated as follows:
$D_i \sim \text{Bernoulli}(0.5)$, $X_{1i} \sim \text{Bernoulli}(0.5)$, $X_{2i} \sim \text{Unif}\{-1,0,1\}$, and $X_{3i} \sim N(0,1)$.
Supposing that the target population shares the same distribution of $(X_1, X_2)$ as the source population, we set $p^\calJ(x) = 1/6$ for all $x \in \calX$.
In addition, we create a mismeasured version of $X_{3i}$ as $X_{3i}^{\text{er}} = X_{3i} + \nu_i$, where $\nu_i \sim \text{Unif}[-0.3,0.3]$.

The network $\bm A^{\calI}$ is generated as follows.
We first draw each unit's degree $G_i$ independently from $\calG = \{0,1,2,3,4\}$.
Then, for each $j \ne i$, we set $A_{ij} = \bm{1}\{\text{dist}_{ij} \le \bar g_i\}$, where $\bar g_i$ is the $G_i$-th smallest element of $\{\text{dist}_{ij}: j \in \calI \setminus \{i\}\}$, and $\text{dist}_{ij}$ is the Mahalanobis distance based on $(X_2, X_3)$.
We also define $\tilde{\text{dist}}_{ij}$ as the Mahalanobis distance based on $(X_2, X_3^{\text{er}})$, which serves as the proxy of $\text{dist}_{ij}$.
The error term follows a network autoregressive process $\epsilon_i = \rho \sum_{j \neq i} \tilde A^{\calI}_{ij} \epsilon_j + u_i$, where $u_i \sim N(0,1)$ and $\tilde A^{\calI}_{ij}$ denotes the $(i,j)$-th element of the row-normalized version of $\bm A^{\calI}$.
The network autoregressive parameter is chosen from $\rho \in \{0.3, 0.5\}$.

To implement our inferential procedure, several functions and parameters need to be specified.  
First, the bandwidth $b = (b_c, b_o)$ in the discrete kernel regression is set as $b = c_b \cdot \hat b_{n^\calI}$, where $c_b$ is a scaling constant chosen from $c_b \in \{0.5, 1, 2\}$, and $\hat b_{n^\calI}$ are optimal bandwidths estimated via leave-one-out cross validation in the kernel regression.\footnote{
    We used the \texttt{npscoef} function in the \texttt{np} package.
    Since performing the cross validation in every iteration is computationally too demanding, we computed the optimal bandwidths for 20 burn-in samples in each setting and used their averages as $\hat b_{n^\calI}$.
}
The solution set $\calS^*_{\delta,q,x}$ is estimated according to \eqref{eq:Shat}, with $a = \hat{\bar{\kappa}}_{\delta,q,x} \cdot (n^\calI)^{-2/5}$.  
To assess the impact of estimating $\calS^*_{\delta,q,x}$ on the precision of inference, we also consider an infeasible estimator that employs the true $\calS^*_{\delta,q,x}$ in line 7 of Algorithm \ref{algo:boot}.  
For the kernel function used in the wild bootstrap, we set $K(u) = \bm{1}\{|u| \le 1\}(1 - u)^2$.  
As a distance measure that combines information of both the covariate distance (which is mismeasured) and network proximity, we consider the following network weighted Mahalanobis distance
\begin{align}
\tilde{\Delta}_{ij} = \gamma_{ij}\tilde{\text{dist}}_{ij}, \;\; \text{where} \;\; \gamma_{ij} = \bm{1}\{j \ne i\}\Phi\left(1 - \frac{1}{\text{path}_{ij} - 1}\right),
\end{align}
and $\text{path}_{ij}$ denotes the shortest-path distance between units $i$ and $j$ on $\bm A^{\calI}$.
For example, $\gamma_{ii} = 0$, $\gamma_{ij} = 0$ if $A^{\calI}_{ij} = 1$ (i.e., $\text{path}_{ij} = 1$), $\gamma_{ij} = 0.5$ if $\text{path}_{ij} = 2$, and so forth.
Defining the distance in this way may be seen as a combination of \cite{kojevnikov2021bootstrap} and \cite{conley2023bootstrap}.
The bandwidth $d$ is set to be the $(c_d \max_{i \in \calI}G_i/n^\calI)$ quantile of $\{\tilde \Delta_{ij}: i,j \in \calI, i \ne j\}$, where $c_d$ is chosen from $c_d \in \{2, 4, 6\}$.
For comparison, we also compute the empirical coverage for the estimator that ignores cross-sectional dependence (i.e., setting $\bbK_\calI = I_{n^\calI}$).

In this analysis, we perform the wild bootstrap to simulate the distribution of $T_{\delta,q} \coloneqq \sqrt{n^\calI}( \hat{\bar \kappa}_{\delta, q} - \bar \kappa_{\delta, q})$, where we consider the Wasserstein ball with $q = 2$ and four radius values $\delta \in \{0.05, 0.1, 0.2, 0.5\}$ centered at the uniform reference distribution $\pi_x^*(g) = 1/5$, for all $g \in \calG$ and $x \in \calX$.
The number of bootstrap replications is set to $B = 500$, and we compute the 95\% and 99\% bootstrap CIs for $T_{\delta,q}$, checking whether it is contained in each case.
This procedure is repeated for 500 Monte Carlo replications to compute the empirical coverage probabilities.

Tables \ref{tab:rho03} and \ref{tab:rho05} report the results for $\rho = 0.3$ and $\rho = 0.5$, respectively.
The main findings are as follows.
When the dependence of the error terms is relatively weak ($\rho = 0.3$), our wild bootstrap method performs well overall.
In particular, when the sample size is large, the empirical coverage rates are satisfactorily close to the nominal levels in almost all cases.
The choice of the two bandwidths, one in the discrete kernel regression and the other in the dependent wild bootstrap, has relatively a small influence on the results.
Moreover, the effect of estimating $\calS^*_{\delta,q,x}$ on the coverage accuracy is almost negligible, which is consistent with our theory.
By contrast, when the network dependence among error terms is ignored, the resulting CIs are clearly too narrow, especially for smaller samples.
As the magnitude of network dependence increases ($\rho = 0.5$), this undercoverage becomes more serious.
Even for our dependent wild bootstrap, a slight loss of coverage is observed for smaller samples, but the accuracy improves as the sample size grows.
Similar results to ours have been reported in previous studies, such as \cite{kim2011spatial}.
Overall, these results confirm that the proposed wild bootstrap procedure performs reliably and is relatively insensitive to the choice of tuning parameters, at least for this particular DGP.
Ideally, a fully data-driven method for selecting these factors could be developed, but we leave this for future research.

\begin{table}[!h]
\footnotesize\centering
\caption{Empirical coverage probabilities: $\rho = 0.3$}\label{tab:rho03}
\begin{tabular}{rrccrrrrrrrr}
\toprule
\multicolumn{4}{c}{} & \multicolumn{4}{c}{95\% CI} & \multicolumn{4}{c}{99\% CI} \\
\cmidrule(l{3pt}r{3pt}){5-8} \cmidrule(l{3pt}r{3pt}){9-12}
$n^\calI$ & $c_b$ & $\calS^*$ & $\bbK$ & $\delta = 0.05$ & $0.1$ & $0.2$ & $0.5$ & $\delta = 0.05$ & $0.1$ & $0.2$ & $0.5$\\
\midrule
400  & 0.5 & est  & $c_d = 2$ & 0.942 & 0.942 & 0.934 & 0.930 & 0.982 & 0.982 & 0.978 & 0.980 \\
     &     &      & $c_d = 4$ & 0.942 & 0.940 & 0.940 & 0.926 & 0.984 & 0.984 & 0.984 & 0.982 \\
     &     &      & $c_d = 6$ & 0.934 & 0.934 & 0.930 & 0.926 & 0.978 & 0.976 & 0.978 & 0.976 \\
     &     &      & $\bbK = I$ & 0.854 & 0.852 & 0.854 & 0.856 & 0.946 & 0.948 & 0.948 & 0.944 \\
     &     & true & $c_d = 2$ & 0.944 & 0.944 & 0.944 & 0.942 & 0.982 & 0.982 & 0.980 & 0.984 \\
     &     &      & $c_d = 4$ & 0.942 & 0.938 & 0.940 & 0.938 & 0.984 & 0.984 & 0.984 & 0.984 \\
     &     &      & $c_d = 6$ & 0.936 & 0.934 & 0.934 & 0.936 & 0.978 & 0.978 & 0.980 & 0.982 \\
     &     &      & $\bbK = I$ & 0.854 & 0.854 & 0.854 & 0.866 & 0.946 & 0.944 & 0.940 & 0.946 \\
     & 1.0 & est  & $c_d = 2$ & 0.948 & 0.946 & 0.938 & 0.940 & 0.978 & 0.978 & 0.982 & 0.986 \\
     &     &      & $c_d = 4$ & 0.954 & 0.954 & 0.942 & 0.948 & 0.986 & 0.986 & 0.990 & 0.984 \\
     &     &      & $c_d = 6$ & 0.942 & 0.942 & 0.940 & 0.938 & 0.984 & 0.984 & 0.978 & 0.982 \\
     &     &      & $\bbK = I$ & 0.862 & 0.862 & 0.860 & 0.856 & 0.950 & 0.946 & 0.942 & 0.954 \\
     &     & true & $c_d = 2$ & 0.948 & 0.948 & 0.946 & 0.950 & 0.978 & 0.978 & 0.980 & 0.984 \\
     &     &      & $c_d = 4$ & 0.952 & 0.952 & 0.948 & 0.956 & 0.986 & 0.986 & 0.986 & 0.986 \\
     &     &      & $c_d = 6$ & 0.942 & 0.942 & 0.942 & 0.946 & 0.984 & 0.984 & 0.984 & 0.988 \\
     &     &      & $\bbK = I$ & 0.862 & 0.864 & 0.862 & 0.866 & 0.952 & 0.952 & 0.952 & 0.950 \\
     & 2.0 & est  & $c_d = 2$ & 0.946 & 0.946 & 0.940 & 0.948 & 0.986 & 0.986 & 0.986 & 0.988 \\
     &     &      & $c_d = 4$ & 0.960 & 0.958 & 0.954 & 0.954 & 0.988 & 0.988 & 0.988 & 0.986 \\
     &     &      & $c_d = 6$ & 0.952 & 0.952 & 0.942 & 0.946 & 0.980 & 0.980 & 0.980 & 0.984 \\
     &     &      & $\bbK = I$ & 0.876 & 0.874 & 0.870 & 0.864 & 0.952 & 0.952 & 0.952 & 0.954 \\
     &     & true & $c_d = 2$ & 0.946 & 0.946 & 0.946 & 0.952 & 0.986 & 0.986 & 0.988 & 0.986 \\
     &     &      & $c_d = 4$ & 0.960 & 0.960 & 0.960 & 0.960 & 0.988 & 0.988 & 0.988 & 0.990 \\
     &     &      & $c_d = 6$ & 0.950 & 0.950 & 0.948 & 0.954 & 0.980 & 0.980 & 0.980 & 0.982 \\
     &     &      & $\bbK = I$ & 0.876 & 0.874 & 0.874 & 0.874 & 0.952 & 0.954 & 0.954 & 0.960 \\
1200 & 0.5 & est  & $c_d = 2$ & 0.954 & 0.954 & 0.952 & 0.956 & 0.992 & 0.992 & 0.992 & 0.988 \\
     &     &      & $c_d = 4$ & 0.960 & 0.960 & 0.956 & 0.954 & 0.994 & 0.994 & 0.994 & 0.988 \\
     &     &      & $c_d = 6$ & 0.964 & 0.962 & 0.958 & 0.962 & 0.992 & 0.992 & 0.992 & 0.990 \\
     &     &      & $\bbK = I$ & 0.890 & 0.890 & 0.878 & 0.884 & 0.962 & 0.962 & 0.950 & 0.954 \\
     &     & true & $c_d = 2$ & 0.954 & 0.954 & 0.956 & 0.964 & 0.992 & 0.992 & 0.992 & 0.990 \\
     &     &      & $c_d = 4$ & 0.960 & 0.960 & 0.960 & 0.960 & 0.994 & 0.994 & 0.994 & 0.990 \\
     &     &      & $c_d = 6$ & 0.966 & 0.966 & 0.966 & 0.964 & 0.992 & 0.992 & 0.992 & 0.990 \\
     &     &      & $\bbK = I$ & 0.890 & 0.888 & 0.888 & 0.892 & 0.962 & 0.962 & 0.960 & 0.964 \\
     & 1.0 & est  & $c_d = 2$ & 0.956 & 0.954 & 0.952 & 0.956 & 0.992 & 0.992 & 0.992 & 0.990 \\
     &     &      & $c_d = 4$ & 0.962 & 0.960 & 0.956 & 0.952 & 0.994 & 0.994 & 0.994 & 0.992 \\
     &     &      & $c_d = 6$ & 0.964 & 0.962 & 0.960 & 0.964 & 0.992 & 0.992 & 0.992 & 0.992 \\
     &     &      & $\bbK = I$ & 0.894 & 0.892 & 0.884 & 0.888 & 0.966 & 0.964 & 0.958 & 0.954 \\
     &     & true & $c_d = 2$ & 0.956 & 0.956 & 0.960 & 0.964 & 0.992 & 0.992 & 0.992 & 0.992 \\
     &     &      & $c_d = 4$ & 0.962 & 0.962 & 0.960 & 0.958 & 0.994 & 0.994 & 0.994 & 0.992 \\
     &     &      & $c_d = 6$ & 0.964 & 0.964 & 0.966 & 0.962 & 0.992 & 0.992 & 0.992 & 0.990 \\
     &     &      & $\bbK = I$ & 0.894 & 0.896 & 0.892 & 0.900 & 0.966 & 0.966 & 0.964 & 0.964 \\
     & 2.0 & est  & $c_d = 2$ & 0.962 & 0.962 & 0.956 & 0.958 & 0.992 & 0.992 & 0.990 & 0.988 \\
     &     &      & $c_d = 4$ & 0.966 & 0.968 & 0.958 & 0.954 & 0.992 & 0.992 & 0.992 & 0.992 \\
     &     &      & $c_d = 6$ & 0.962 & 0.962 & 0.962 & 0.956 & 0.992 & 0.992 & 0.992 & 0.990 \\
     &     &      & $\bbK = I$ & 0.898 & 0.900 & 0.890 & 0.890 & 0.970 & 0.970 & 0.966 & 0.968 \\
     &     & true & $c_d = 2$ & 0.962 & 0.962 & 0.960 & 0.966 & 0.992 & 0.992 & 0.992 & 0.994 \\
     &     &      & $c_d = 4$ & 0.966 & 0.968 & 0.970 & 0.964 & 0.992 & 0.994 & 0.994 & 0.994 \\
     &     &      & $c_d = 6$ & 0.962 & 0.962 & 0.962 & 0.964 & 0.994 & 0.992 & 0.992 & 0.992 \\
     &     &      & $\bbK = I$ & 0.900 & 0.902 & 0.898 & 0.902 & 0.970 & 0.970 & 0.970 & 0.968 \\
\bottomrule
\end{tabular}
\flushleft

NOTE: "est" and "true" in the column $\calS^*$ indicate that the estimated and true $\calS^*_{\delta,q,x}$ are used, respectively.
In the column $\bbK$, "$\bbK = I$" indicates that network dependence is ignored in this case.
\normalsize
\end{table}

\begin{table}[!h]
\footnotesize\centering
\caption{Empirical coverage probabilities: $\rho = 0.5$}\label{tab:rho05}
\begin{tabular}{rrccrrrrrrrr}
\toprule
\multicolumn{4}{c}{} & \multicolumn{4}{c}{95\% CI} & \multicolumn{4}{c}{99\% CI} \\
\cmidrule(l{3pt}r{3pt}){5-8} \cmidrule(l{3pt}r{3pt}){9-12}
$n^\calI$ & $c_b$ & $\calS^*$ & Estimator & $\delta = 0.05$ & $0.1$ & $0.2$ & $0.5$ & $\delta = 0.05$ & $0.1$ & $0.2$ & $0.5$\\
\midrule
400  & 0.5 & est  & $c_d = 2$ & 0.934 & 0.932 & 0.922 & 0.918 & 0.976 & 0.976 & 0.974 & 0.978 \\
     &     &      & $c_d = 4$ & 0.942 & 0.940 & 0.930 & 0.924 & 0.984 & 0.984 & 0.980 & 0.978 \\
     &     &      & $c_d = 6$ & 0.930 & 0.930 & 0.914 & 0.912 & 0.974 & 0.974 & 0.972 & 0.972 \\
     &     &      & $\bbK = I$ & 0.804 & 0.802 & 0.788 & 0.788 & 0.902 & 0.896 & 0.892 & 0.902 \\
     &     & true & $c_d = 2$ & 0.934 & 0.934 & 0.932 & 0.938 & 0.976 & 0.976 & 0.976 & 0.976 \\
     &     &      & $c_d = 4$ & 0.942 & 0.942 & 0.944 & 0.938 & 0.984 & 0.984 & 0.982 & 0.980 \\
     &     &      & $c_d = 6$ & 0.930 & 0.930 & 0.930 & 0.924 & 0.974 & 0.974 & 0.974 & 0.976 \\
     &     &      & $\bbK = I$ & 0.804 & 0.802 & 0.804 & 0.802 & 0.902 & 0.902 & 0.904 & 0.916 \\
     & 1.0 & est  & $c_d = 2$ & 0.940 & 0.934 & 0.930 & 0.932 & 0.976 & 0.976 & 0.974 & 0.978 \\
     &     &      & $c_d = 4$ & 0.946 & 0.940 & 0.930 & 0.934 & 0.984 & 0.984 & 0.982 & 0.978 \\
     &     &      & $c_d = 6$ & 0.932 & 0.930 & 0.922 & 0.920 & 0.978 & 0.976 & 0.978 & 0.974 \\
     &     &      & $\bbK = I$ & 0.810 & 0.808 & 0.804 & 0.788 & 0.904 & 0.900 & 0.896 & 0.906 \\
     &     & true & $c_d = 2$ & 0.940 & 0.940 & 0.936 & 0.946 & 0.976 & 0.976 & 0.974 & 0.976 \\
     &     &      & $c_d = 4$ & 0.944 & 0.946 & 0.944 & 0.940 & 0.984 & 0.984 & 0.984 & 0.980 \\
     &     &      & $c_d = 6$ & 0.932 & 0.932 & 0.934 & 0.936 & 0.978 & 0.978 & 0.978 & 0.980 \\
     &     &      & $\bbK = I$ & 0.810 & 0.812 & 0.816 & 0.804 & 0.904 & 0.904 & 0.906 & 0.920 \\
     & 2.0 & est  & $c_d = 2$ & 0.942 & 0.944 & 0.936 & 0.938 & 0.982 & 0.982 & 0.980 & 0.978 \\
     &     &      & $c_d = 4$ & 0.946 & 0.944 & 0.940 & 0.936 & 0.984 & 0.986 & 0.988 & 0.988 \\
     &     &      & $c_d = 6$ & 0.940 & 0.940 & 0.934 & 0.928 & 0.980 & 0.980 & 0.980 & 0.980 \\
     &     &      & $\bbK = I$ & 0.830 & 0.826 & 0.818 & 0.808 & 0.922 & 0.922 & 0.912 & 0.914 \\
     &     & true & $c_d = 2$ & 0.942 & 0.944 & 0.942 & 0.950 & 0.982 & 0.982 & 0.982 & 0.984 \\
     &     &      & $c_d = 4$ & 0.946 & 0.948 & 0.948 & 0.948 & 0.984 & 0.986 & 0.988 & 0.988 \\
     &     &      & $c_d = 6$ & 0.940 & 0.940 & 0.938 & 0.946 & 0.980 & 0.980 & 0.980 & 0.978 \\
     &     &      & $\bbK = I$ & 0.830 & 0.830 & 0.824 & 0.824 & 0.922 & 0.922 & 0.922 & 0.922 \\
1200 & 0.5 & est  & $c_d = 2$ & 0.946 & 0.944 & 0.942 & 0.944 & 0.988 & 0.988 & 0.988 & 0.990 \\
     &     &      & $c_d = 4$ & 0.952 & 0.952 & 0.946 & 0.942 & 0.990 & 0.990 & 0.990 & 0.988 \\
     &     &      & $c_d = 6$ & 0.958 & 0.956 & 0.948 & 0.952 & 0.994 & 0.994 & 0.992 & 0.986 \\
     &     &      & $\bbK = I$ & 0.844 & 0.836 & 0.834 & 0.822 & 0.932 & 0.932 & 0.920 & 0.930 \\
     &     & true & $c_d = 2$ & 0.952 & 0.948 & 0.946 & 0.958 & 0.988 & 0.988 & 0.988 & 0.988 \\
     &     &      & $c_d = 4$ & 0.952 & 0.952 & 0.952 & 0.958 & 0.990 & 0.990 & 0.990 & 0.988 \\
     &     &      & $c_d = 6$ & 0.958 & 0.956 & 0.958 & 0.960 & 0.994 & 0.994 & 0.994 & 0.986 \\
     &     &      & $\bbK = I$ & 0.844 & 0.844 & 0.844 & 0.828 & 0.934 & 0.934 & 0.940 & 0.938 \\
     & 1.0 & est  & $c_d = 2$ & 0.950 & 0.946 & 0.940 & 0.942 & 0.988 & 0.988 & 0.988 & 0.990 \\
     &     &      & $c_d = 4$ & 0.950 & 0.950 & 0.948 & 0.944 & 0.990 & 0.990 & 0.990 & 0.990 \\
     &     &      & $c_d = 6$ & 0.962 & 0.960 & 0.954 & 0.950 & 0.994 & 0.994 & 0.992 & 0.986 \\
     &     &      & $\bbK = I$ & 0.850 & 0.848 & 0.840 & 0.832 & 0.936 & 0.936 & 0.928 & 0.928 \\
     &     & true & $c_d = 2$ & 0.952 & 0.952 & 0.952 & 0.960 & 0.988 & 0.988 & 0.988 & 0.988 \\
     &     &      & $c_d = 4$ & 0.952 & 0.954 & 0.956 & 0.960 & 0.990 & 0.990 & 0.990 & 0.990 \\
     &     &      & $c_d = 6$ & 0.962 & 0.960 & 0.960 & 0.964 & 0.994 & 0.994 & 0.994 & 0.986 \\
     &     &      & $\bbK = I$ & 0.850 & 0.848 & 0.842 & 0.840 & 0.936 & 0.938 & 0.944 & 0.936 \\
     & 2.0 & est  & $c_d = 2$ & 0.960 & 0.958 & 0.946 & 0.944 & 0.988 & 0.988 & 0.988 & 0.988 \\
     &     &      & $c_d = 4$ & 0.954 & 0.954 & 0.954 & 0.948 & 0.992 & 0.990 & 0.990 & 0.990 \\
     &     &      & $c_d = 6$ & 0.960 & 0.960 & 0.958 & 0.952 & 0.994 & 0.994 & 0.994 & 0.990 \\
     &     &      & $\bbK = I$ & 0.858 & 0.860 & 0.848 & 0.846 & 0.940 & 0.938 & 0.932 & 0.930 \\
     &     & true & $c_d = 2$ & 0.960 & 0.960 & 0.954 & 0.962 & 0.988 & 0.988 & 0.988 & 0.988 \\
     &     &      & $c_d = 4$ & 0.954 & 0.954 & 0.958 & 0.960 & 0.992 & 0.992 & 0.992 & 0.990 \\
     &     &      & $c_d = 6$ & 0.960 & 0.960 & 0.962 & 0.962 & 0.994 & 0.994 & 0.994 & 0.990 \\
     &     &      & $\bbK = I$ & 0.858 & 0.856 & 0.850 & 0.844 & 0.940 & 0.940 & 0.944 & 0.934 \\
\bottomrule
\end{tabular}

\flushleft
NOTE: "est" and "true" in the column $\calS^*$ indicate that the estimated and true $\calS^*_{\delta,q,x}$ are used, respectively.
In the column $\bbK$, "$\bbK = I$" indicates that network dependence is ignored in this case.
\normalsize
\end{table}

\section{An Empirical Illustration}\label{sec:empiric}

As an empirical illustration, we apply our bound estimator and wild bootstrap method to the data on farmers' insurance adoption in \cite{cai2015social}.  
\cite{cai2015social} conducted a field experiment to estimate the effect of providing intensive information sessions about the weather insurance on farmers' insurance take-up decisions.  
In the experiment, four types of sessions were provided: first round simple, first round intensive, second round simple, and second round intensive.  
In each round, the simple sessions only explain the insurance contract, while intensive sessions cover all information provided in simple sessions and additionally provide financial education to help farmers understand how the insurance works and its benefits.  
The farmers were randomly assigned to each session according to household size and area of rice production per capita, which we denote by \textit{hhsize} and \textit{rice}, respectively.

In this analysis, the outcome variable is $Y_i \in \{ 0, 1 \}$, which indicates whether farmer $i$ decided to buy the weather insurance after attending the session.
Let $\textit{int}_i \in \{ 0, 1 \}$ denote whether $i$ was assigned to an intensive session, and $\textit{first}_i \in \{ 0, 1 \}$ denote whether $i$ was assigned to the first round session.
The spillover effects matter only for the second round participants, as they can receive information from the first round participants.
Then, as own treatment indicator, we set $D_i = \textit{int}_i$.
Meanwhile, reflecting the experimental design, the exposure variable is defined as follows:
\begin{align}
    E_i = (1 - \textit{first}_i) \sum_{j \in \calI} A_{ij}^\calI \, \textit{int}_j \, \textit{first}_j / G_i,
\end{align}
where $A_{ij}^\calI$ indicates whether $i$ and $j$ are mutual information-exchange partners.
The TTE in this context is given by $\tau_i = Y_i(1,1) - Y_i(0,0)$.
$\tau_i$ interpreted as the individual policy effect for the policy that provides intensive session for all farmers, and they all have enough time to exchange their information with their partners.

As the covariates, we use
$X_{1i} = \bm{1}\left\{ \textit{hhsize}_i \ge \text{Med}[\textit{hhsize}_i] \right\}$ and
$X_{2i} = \bm{1}\left\{ \textit{rice}_i \ge \text{Med}[\textit{rice}_i] \right\}$, where $\text{Med}$ denotes the empirical median.
For the basis function $w$, we consider the following form: $w(d,e,g) = (1, d, e, de, \log(g + 1), e\log(g + 1))$.

To evaluate the performance of our proposed method in a realistic setting, we randomly divide the original data into two groups.
Specifically, since the data consist of 47 administrative villages, we randomly select 17 villages as the source sample ($n^\calI = 1514$) and use the remaining 30 villages as the target sample ($n^\calJ = 3351$).
We then compute the ATTE bounds for the target sample by transferring the estimates obtained from the source sample.
In this analysis, because the network structure in the target data is actually known, we can directly compute $\pi^\calJ(g,x)$ for each $(g,x)$.
This enables us to approximately assess the coverage property of our bound estimator under different choices of the Wasserstein radius $\delta$.
In addition, to illustrate the effect of increasing the size of source sample, we also consider a case in which five additional villages are included in the source sample ($n^\calI = 1812$).

\begin{figure}[ht]
\begin{center}
    \includegraphics[width=16cm]{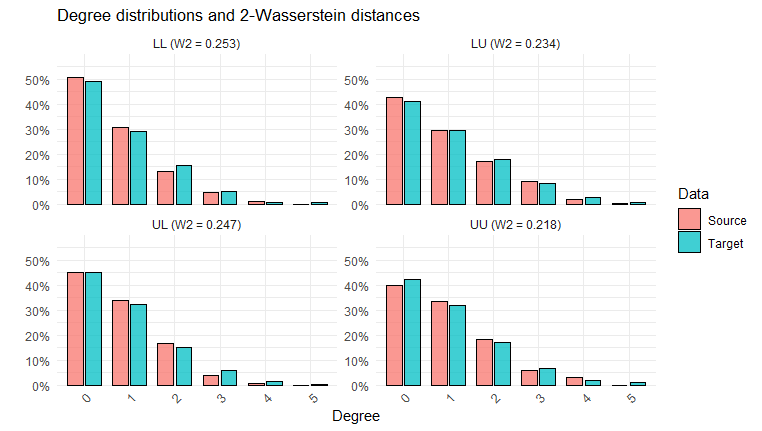}
    \caption{Conditional degree distributions}
    \label{fig:cond_deg}
\end{center}
\end{figure}

To determine a plausible range for $\delta$, we compute the 2-Wasserstein distance between the degree distributions of the 17 source villages and 30 target villages for each covariate group (throughout this analysis, we use the 2-Wasserstein distance).
The results are reported in Figure \ref{fig:cond_deg}.
In the figure, "LL" stands for the subsample with $(X_1 = 0, X_2 = 0)$, "LU" for $(X_1 = 0, X_2 = 1)$, and so on.
The number shown at the top of each panel indicates the computed 2-Wasserstein distance.
From these results, we observe that when the source and target data are drawn from the same population, the typical 2-Wasserstein distance is roughly 0.25 or so.

Based on the above finding, we slightly conservatively set the region for $\delta$ as $\delta \in (0, 0.6]$.
As the baseline conditional degree distribution, we set $\pi_x^*(g) = \pi^\calI(g, x)$ (see Figure \ref{fig:cond_deg}).
The distance measure $\tilde \Delta_{ij}$ is computed using the Mahalanobis distance based on age, gender, acreage of rice production, and household size, weighted by the path length as in Section \ref{sec:MC}. 
Furthermore, when $i$ and $j$ belong to different villages, we set $\tilde{\Delta}_{ij} = \infty$.
All other setups for estimation and bootstrap inference follow those used in the simulation analysis in Section \ref{sec:MC}.

\begin{figure}[ht]
    \caption{Estimated coefficient functions}\label{fig:sensitivity}
    \centering
    \begin{subfigure}[b]{0.48\textwidth}
        \centering
        \includegraphics[width = 9cm]{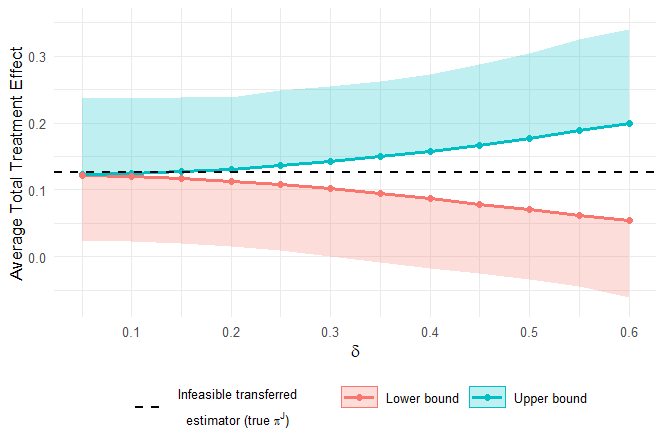}
        \caption{Sensitivity analysis result (17 villages: $n^\calI = 1514$)}
        \label{fig:result17}
    \end{subfigure}
    \begin{subfigure}[b]{0.48\textwidth}
        \centering
        \includegraphics[width = 9cm]{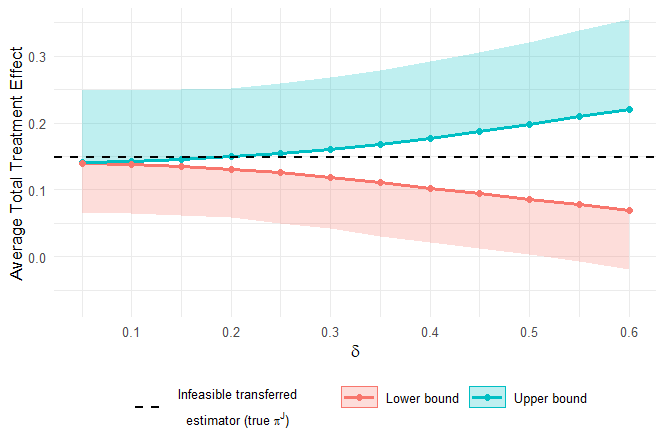}
        \caption{Sensitivity analysis result (22 villages: $n^\calI = 1812$)}
        \label{fig:result22}
    \end{subfigure}
\end{figure}

We report our bound estimation results in Figure \ref{fig:sensitivity}.
In the figure, the left and right panels correspond to the cases with 17 and 22 villages in the source sample, respectively.
The upper shaded area represents the upper half of the 95\% CI for the upper bound, and the lower shaded area shows the lower half of the 95\% CI for the lower bound.
The dashed line indicates the (infeasible) point estimate of $\kappa^\calJ$, computed using the true conditional degree distribution $\pi^\calJ$ in the target data.
Since the source and target datasets come from essentially the same population and $\pi^\calI \approx \pi^\calJ$ holds, as shown in Figure \ref{fig:cond_deg}, our ATTE bound is highly informative, successfully covering the estimated $\kappa^\calJ$ even for small values of $\delta$.
In addition, the estimated worst case bound does not fall below zero for any $\delta \le 0.6$ for both sample sizes.
Regarding the impact of increasing the size of the source sample, we can observe that the length of the CI for each $\delta$ is significantly narrower in the right panel than in the left.
When 22 villages are used for the source sample, the lower 95\% bound remains positive for almost the entire range of $\delta$ values considered here.
From these findings, we may state that the ATTE is likely positive for the target data with a certain degree of confidence.

\section{Conclusion}\label{sec:conclusion}

This paper proposes a transfer learning framework for policy evaluation in settings where the network structure of the target data is unobserved.
Following the existing literature, we adopt a covariate-shift type assumption to estimate conditional mean potential outcomes using experimental source data.
However, in the presence of spillover effects, this assumption alone is insufficient to evaluate a specific policy in the target data due to the lack of network information.
To address this issue, we propose a sensitivity analysis approach that quantifies the uncertainty in the unobserved target network using the Wasserstein distance between degree distributions.
The resulting bounds on the policy effect can be computed by solving a set of linear programming problems.
We derive the asymptotic distribution of the bound estimator via the functional delta method and develop a wild bootstrap procedure for inference.
As an empirical application, we use the experimental data from \cite{cai2015social} to illustrate the practical implementation and empirical usefulness of the proposed method.

Several limitations should be noted.
First, the covariate-shift assumption may be violated if the source and target data are too dissimilar.
Second, the current model specification assumes that network effects can be entirely captured by node-level covariates, not allowing any network-level heterogeneity.
Third, the proposed framework cannot be directly applied to the evaluation of more complex policies that assign treatment based on individual characteristics or network positions.
Finally, as with any sensitivity analysis, the interpretation and selection of the uncertainty parameter ($\delta$ in our context) remain open questions.

\section*{Acknowledgments}
I thank Hanbat Jeong, Kensuke Sakamoto, and the participants of seminars at the University of Melbourne and Macquarie University for their valuable comments and constructive suggestions.
This work was supported by JSPS KAKENHI Grant Number 23KK0226.
Most parts of this paper were written during my research visit at the Melbourne Business School (MBS).
I am deeply grateful to MBS for their hospitality.

\clearpage
\appendix
\begin{center}
    \Large \textbf{Appendix}
\end{center}
\section{Technical Appendix}

\subsection{Derivation of the dual problem (\ref{eq:dual})}\label{app:dual}

Recall that our primal linear program is formulated as follows:
\begin{align}
    &\text{maximize } \sum_{x \in \calX} \sum_{u,v \in \calG^2} \Gamma_x(u,v) m(v, x) \\
    &\text{subject to } \sum_{v \in \calG}\Gamma_x(u, v) = \pi_x^*(u), \sum_{u, v \in \calG^2} \Gamma_x(u, v) \bigl|u - v\bigr|^q \le \delta^q, \Gamma_x(u, v) \ge 0, \forall\, (x, u, v) \in \calX \times \calG^2 
\end{align}
Now, introduce dual variables $n(u, x)$ for the equality constraint $\sum_{v \in \calG}\Gamma_x(u, v) = \pi_x^*(u)$ for each $(x, u) \in \calX \times \calG$ and $\lambda_x \ge 0$ for the inequality constraint $\sum_{u, v \in \calG^2} \Gamma_x(u, v) \bigl|u - v\bigr|^q \le \delta^q$ for each $x \in \calX$.
Then, the Lagrangian function is given by
\begin{align}
    L(\Gamma, n,\lambda)
    & = \sum_{x \in \calX} \sum_{u,v \in \calG^2} \Gamma_x(u,v) m(v, x) - \sum_{x \in \calX} 
 \sum_{u \in \calG}n(u,x) \left(\sum_{v \in \calG}\Gamma_x(u, v) - \pi_x^*(u)\right) \\
    & \quad - \sum_{x \in \calX} \lambda_x \left( \sum_{u, v \in \calG^2} \Gamma_x(u, v) \bigl|u - v\bigr|^q - \delta^q \right) \\
    & = \sum_{x \in \calX} \lambda_x \delta^q + \sum_{x \in \calX}\sum_{u \in \calG}n(u, x)\pi_x^*(u) + \sum_{x \in \calX} \sum_{u,v \in \calG^2}\Gamma_x(u,v) \left\{m(v, x) - n(u,x)-\lambda_x|u-v|^{q}\right\}.
\end{align}
Define the dual function by
\begin{align}
    D(n, \lambda) 
    & \coloneqq \sup_{\Gamma \ge 0}L(\Gamma, n,\lambda) \\
    & =  \sum_{x \in \calX} \lambda_x \delta^q + \sum_{x \in \calX}\sum_{u \in \calG}n(u, x)\pi_x^*(u) + \sup_{\Gamma \ge 0}\sum_{x \in \calX} \sum_{u,v \in \calG^2}\Gamma_x(u,v) \left\{m(v, x) - n(u,x)-\lambda_x|u-v|^{q}\right\}.
\end{align}
If the following inequality is not satisfied
\begin{align}\label{eq:const}
    m(v,x)-n(u,x)-\lambda_x|u-v|^q \le 0 
\end{align}
for some $(x,u,v)$, then we can set the corresponding element of $\Gamma_x(u,v)$ arbitrarily large, resulting in an unbounded $D(n, \lambda)$.
Thus, whenever \eqref{eq:const} is satisfied, we must have
\begin{align}
    D(n,\lambda) = \sum_{x \in \calX} \left( \lambda_x \delta^q + \sum_{u \in \calG} n(u,x)\pi^*_x(u) \right).
\end{align}
To minimize the dual function $D(n,\lambda)$, in view of \eqref{eq:const}, we can profile out $n$ from $D(n,\lambda)$ by setting
\begin{align}
    n(u, x)= \max_{v \in \calG}\{m(v, x) - \lambda_x |u-v|^q\}.
\end{align}
Plugging this into $\lambda_x \delta^q + \sum_{u \in \calG} n(u,x)\pi^*_x(u)$ gives the objective function in \eqref{eq:dual}.

\subsection{The dual problem of Example \ref{exp:simple}}\label{app:exp}

The dual problem of Example \ref{exp:simple} is as follows: $\min_{\lambda \ge 0} D(\lambda)$, where
\begin{align}
     D(\lambda) \coloneqq \left\{ \lambda \delta + \sum_{u \in \calG} \left[ \max_{v \in \calG}\{ m(v) - \lambda |u - v|\} \right] \pi^*(u) \right\}.
\end{align}    
By direct calculation,
\begin{align}
\sum_{u \in \calG} \left[ \max_{v \in \calG}\{ m(v) - \lambda |u - v|\} \right] \pi^*(u) 
& = \max\{ m(0) - \lambda, m(1)\}\alpha^* + \max\{ m(0), m(1) - \lambda\}(1 - \alpha^*) \\
& = m(1) \alpha^* + \max\{ m(0), m(1) - \lambda\}(1 - \alpha^*).
\end{align}
Now, when $\lambda > m(1) - m(0)$, 
\begin{align}
    \min_{\lambda > m(1) - m(0)}D(\lambda)
    & = \min_{\lambda > m(1) - m(0)}\left\{ \lambda \delta + m(1) \alpha^* + m(0)(1 - \alpha^*) \right\} \\
    & > m(0) + (m(1) - m(0)) (\alpha^* + \delta).
\end{align}
Meanwhile, if $\lambda \le m(1) - m(0)$,
\begin{align}
    \min_{0 \le \lambda \le m(1) - m(0)}D(\lambda)
    & = \min_{0 \le \lambda \le m(1) - m(0)}\left\{ \lambda \delta + m(1) - \lambda (1 - \alpha^*) \right\}.
\end{align}
Hence, if $\delta \ge (1 - \alpha^*)$, we should set $\lambda =0$, leading to $\min_{0 \le \lambda \le m(1) - m(0)}D(\lambda) = m(1)$.
On the other hand, if $\delta < (1 - \alpha^*)$, the optimal $\lambda$ is given by $m(1) - m(0)$, leading to $\min_{0 \le \lambda \le m(1) - m(0)}D(\lambda) = m(0) + (m(1) - m(0)) (\alpha^* + \delta)$.
Then, it is straightforward to see that $\min_{\lambda \ge 0} D(\lambda) = \overline{\kappa}_{\delta,1}$ holds.

\subsection{Proofs of Proposition \ref{prop:m}, Theorem \ref{thm:dist}, and Theorem \ref{thm:boot}}\label{app:m}

Throughout the proofs, we use the following notations: 
\begin{align}
    \Sigma_{n^\calI}(x) 
    & \coloneqq \frac{1}{n^{\calI}} \sum_{i \in \calI} \bbE[ W_i W_i^\top \bm{1}\{X_i = x\}] \\
    \Omega_{n^\calI}(x)
    & \coloneqq  \frac{1}{n^{\calI}} \sum_{i, i' \in \calI} \bbE\left[ W_i W_{i'}^\top \epsilon_i \epsilon_{i'} \bm{1}\{X_i = X_{i'} = x\} \right] \\
    \Omega_{n^\calI}(x_1, x_2)
    & \coloneqq  \frac{1}{n^{\calI}} \sum_{i, i' \in \calI} \bbE\left[ W_i W_{i'}^\top \epsilon_i \epsilon_{i'} \bm{1}\{X_i = x_1, X_{i'} = x_2\} \right]
\end{align}

\begin{align}
    \hat{\bm{J}}_{n^\calI}
    & \coloneqq \left( \begin{array}{cccc}
    \left( \frac{1}{n^\calI} \sum_{i \in \calI} W_i W_i^\top L_{i,b}( x_1) \right)^{-1} & \bm{0}_{d_w \times d_w} & \cdots & \bm{0}_{d_w \times d_w} \\
    \bm{0}_{d_w \times d_w} & \left( \frac{1}{n^\calI} \sum_{i \in \calI} W_i W_i^\top L_{i,b}(x_2) \right)^{-1}  & \cdots & \bm{0}_{d_w \times d_w} \\
    \vdots & \vdots  & \ddots & \vdots \\
    \bm{0}_{d_w \times d_w} & \bm{0}_{d_w \times d_w}  & \cdots & \left( \frac{1}{n^\calI} \sum_{i \in \calI} W_i W_i^\top L_{i,b}(x_{d_x}) \right)^{-1} 
    \end{array} \right) \\
    \bm{J}_{n^\calI}
    & \coloneqq \left( \begin{array}{cccc}
    \left( \Sigma_{n^\calI}(x_1) \right)^{-1} & \bm{0}_{d_w \times d_w} & \cdots & \bm{0}_{d_w \times d_w} \\
    \bm{0}_{d_w \times d_w} & \left( \Sigma_{n^\calI}(x_2) \right)^{-1}  & \cdots & \bm{0}_{d_w \times d_w} \\
    \vdots & \vdots  & \ddots & \vdots \\
    \bm{0}_{d_w \times d_w} & \bm{0}_{d_w \times d_w}  & \cdots & \left( \Sigma_{n^\calI}(x_{d_x}) \right)^{-1}
    \end{array} \right) \\
    \bm{\Omega}_{n^\calI} 
    & \coloneqq \left( \begin{array}{cccc}
    \Omega_{n^\calI}(x_1) & \Omega_{n^\calI}(x_1, x_2) & \cdots & \Omega_{n^\calI}(x_1, x_{d_x}) \\
     \Omega_{n^\calI}(x_2, x_1) & \Omega_{n^\calI}(x_2) & \cdots & \Omega_{n^\calI}(x_2, x_{d_x}) \\
     \vdots & \vdots & \ddots & \vdots \\
     \Omega_{n^\calI}(x_{d_x}, x_1) & \Omega_{n^\calI}(x_{d_x}, x_2) & \cdots & \Omega_{n^\calI}(x_{d_x}) 
    \end{array} \right)
\end{align}
and
\begin{align}
    \underbracket{Z(x)}_{d_g \times d_w} 
    \coloneqq \left( \begin{array}{c}
    z(g_1, x)^\top \\
    z(g_2, x)^\top \\
     \vdots \\
    z(g_{d_g}, x)^\top
    \end{array} \right), \quad
    \underbracket{\bm{Z}}_{d_g d_x \times d_w d_x} 
    \coloneqq \left( \begin{array}{cccc}
     Z(x_1) & \bm{0}_{d_g \times d_w} & \cdots & \bm{0}_{d_g \times d_w} \\
    \bm{0}_{d_g \times d_w} & Z(x_2)  & \cdots & \bm{0}_{d_g \times d_w} \\
    \vdots & \vdots  & \ddots & \vdots \\
    \bm{0}_{d_g \times d_w} & \bm{0}_{d_g \times d_w}  & \cdots & Z(x_{d_x})
    \end{array} \right).
\end{align}
Moreover, we write $\bm{\beta} = (\beta(x_1)^\top, \beta(x_2)^\top, \ldots, \beta(x_{d_x})^\top)^\top$, $\hat{\bm{\beta}} = (\hat \beta(x_1)^\top, \hat \beta(x_2)^\top, \ldots, \hat \beta(x_{d_x})^\top)^\top$, $m(x) \coloneqq (m(g_1, x), \ldots, m(g_{d_g}, x))^\top$, and $\bm{m} \coloneqq (m(x_1)^\top, \ldots, m(x_{d_x})^\top)^\top$.

\begin{assumption}\label{as:vc}
    \begin{enumerate}
        \item $||w(d,e,g)|| \le c_w < \infty$ a.s. uniformly in $(d,e,g) \in \{0,1\} \times \calE \times \calG$.
        \item For all $i \in \calI$, 
        \begin{align}
            \epsilon_i = \sum_{j \in \calI} r_{ij} \varepsilon_j,
        \end{align}
        where $r_{ij}$ is a non-stochastic possibly unknown weight; $\{\varepsilon_i\}$ are independently and identically distributed over $\calI$, independent of $\{(W_i, X_i)\}$, with mean zero and variance $\sigma_\varepsilon^2$; 
        $\bbE|\varepsilon_i|^4 < \infty$; 
        and $\max\{ \max_{i \in \calI}\sum_{j \in \calI}|r_{ij}|, \max_{j \in \calI}\sum_{i \in \calI}|r_{ij}|\} \le c_r < \infty$, uniformly in $n^\calI$.
        \item  $\Sigma_{n^\calI}(x)$, $\Omega_{n^\calI}(x)$, and $\bm{\Omega}_{n^\calI}$ are positive definite for all sufficiently large $n^\calI$.
        \item For all $x \in \calX$,
        \begin{align}
            \left\| \frac{1}{n^\calI} \sum_{i \in \calI} W_i W_i^\top \bm{1}\{X_i = x\} - \Sigma_{n^\calI}(x) \right\| = O_P\left( 1/\sqrt{n^\calI} \right)
        \end{align}
        \item There exists $b \in (0,1)$ such that $(b_c, b_o) \asymp b$ and $\sqrt{n^\calI} b \to 0$.
    \end{enumerate}
\end{assumption}

Assumption \ref{as:vc}.1 is standard in applications.  
\ref{as:vc}.2 allows for cross-sectional dependence in the error terms.  
For example, if $r_{ij} = A^\calI_{ij}$, the last condition implies that each individual has only finitely many interacting partners.  
The same type of error structure has often been considered in the literature (e.g., \citealp{kelejian2007hac, conley2023bootstrap}).  
\ref{as:vc}.3 is a standard non-singularity condition.  
It also requires that the proportion of each $x$-value is nondegenerate.
\ref{as:vc}.4 is high-level but can be satisfied under appropriate weak dependence conditions on $\{(W_i, X_i)\}$.
Finally, \ref{as:vc}.5 is a technical condition to eliminate the bias in the kernel regression.

\bigskip

Next, we introduce assumptions used to establish the validity of the wild bootstrap procedure.
Let
\begin{align}
    \calB_{i,\calI} \coloneq \{j \in \calI: \tilde \Delta_{ij} \le d\}, \quad \lambda_{i, \calI} \coloneqq |\calB_{i,\calI}|, \quad \lambda_{\calI} \coloneqq \frac{1}{n^\calI} \sum_{i \in \calI} \lambda_{i, \calI}, \quad V_i \coloneqq \left(
    \begin{array}{c}
     W_i \bm{1}\{X_i = x_1\} \\
     W_i \bm{1}\{X_i = x_2\} \\
    \vdots \\
     W_i \bm{1}\{X_i = x_{d_x}\}
    \end{array}
     \right) \epsilon_i,
\end{align}
where $d$ is the bandwidth parameter used in the kernel function $K$.
\begin{assumption}\label{as:boot}
    \begin{enumerate}
        \item $K(s) = K(-s)$ for all $s \in \bbR$ and $K(0) = 1$; $ \sup_{i\in\calI} \bbE(\sum_{j \notin \calB_{i,\calI}} |K(\tilde \Delta_{ij}/d)|) / \bbE \lambda_{\calI} = O(1)$; $ \sup_{i\in\calI} \sum_{j \notin \calB_{i,\calI}} |K(\tilde \Delta_{ij}/d)| / \bbE \lambda_{\calI} = O_P(1)$; $\bbK_\calI$ is symmetric and positive semidefinite a.s.
        \item There exists $c_{q_0} > 0$ such that $(n^\calI)^{-1}\sum_{i,j \in \calI} ||\bbE[V_i V_j^\top]||\Delta_{ij}^{q_0} < c_{q_0}$, where $q_0$ denotes the Parzen characteristic exponent of the kernel function $K$.
        \item $\{\nu_{ij}\}$ are independent of $\{(W_i, X_i, \varepsilon_i)\}$ and are uniformly bounded in $i, j \in \calI$.
        \item For all $i \in \calI$, $\lambda_{i, \calI} \le c \bbE \lambda_\calI$ a.s., for some $c > 0$.
        \item $d \to \infty$, and $\bbE \lambda_\calI \to \infty$ such that $\bbE \lambda_\calI / \sqrt{n^\calI} \to 0$.
        \item $a \downarrow 0$ such that $\sqrt{n^\calI} a \to \infty$.
    \end{enumerate}
\end{assumption}

Assumptions \ref{as:boot}.1, \ref{as:boot}.2, \ref{as:boot}.3, and \ref{as:boot}.4 correspond, respectively, to Assumptions 1, 3, 4, and 5 in \cite{conley2023bootstrap}.  
Specifically, \ref{as:boot}.1 collects the conditions on the kernel weight function.  
As noted in Footnote \ref{foot:psd}, the positive semidefinite-ness of $\bbK_\calI$ is a high-level condition.  
\cite{conley2023bootstrap} provide an alternative bootstrap procedure for situations where this condition fails.  
\ref{as:boot}.2 requires that the dependence between $i$ and $j$ decays as the true distance $\Delta_{ij}$ increases.  
The formal definition of the Parzen characteristic exponent $q_0$, along with related discussion, can be found for example in \cite{andrews1991heteroskedasticity} and \cite{conley2023bootstrap}. 
\ref{as:boot}.3 requires that the measurement errors are independent and uniformly bounded, which is standard in the HAC estimation literature.  
\ref{as:boot}.4 restricts the number of neighbors each unit can have to be of the same order. 
\ref{as:boot}.5 imposes conditions on the bandwidth $d$.
Finally, \ref{as:boot}.6 is a technical condition needed to ensure the consistency of $\hat{\calS}^*_{\delta,q,x}$ for $\calS^*_{\delta,q,x}$.


\begin{flushleft}
    \textbf{Proof of Proposition \ref{prop:m}}
\end{flushleft}

(i) Let $\ell_{i,b}(x) \coloneqq L_{i,b}(x) - \bm{1}\{X_i = x\}$.
It is easy to see that $\ell_{i,b}(x) \le c \cdot b$.
To see this, for example, suppose that $d_o = d_c = 1$.
Then,
\begin{align}
    \ell_{i,b}(x) 
    & = \bm{1}\{X_i^c \neq x^c, X_i^o = x^o\} b_c + \bm{1}\{X_i^c = x^c, X_i^o \neq x^o\} b_o^{|X_i^o - x^o|} + \bm{1}\{X_i^c \neq x^c, X_i^o \neq x^o\} b_c b_o^{|X_i^o - x^o|}.
\end{align}
With this and Assumptions \ref{as:vc}.1 and  \ref{as:vc}.4,
\begin{align}\label{eq:Sigma}
    \begin{split}
     \frac{1}{n^\calI} \sum_{i \in \calI} W_i W_i^\top L_{i,b}(x) 
     & =  \frac{1}{n^\calI} \sum_{i \in \calI} W_i W_i^\top \bm{1}\{X_i = x\} +  \frac{1}{n^\calI} \sum_{i \in \calI} W_i W_i^\top \ell_{i,b}(x) \\
     & = \frac{1}{n^\calI} \sum_{i \in \calI} W_i W_i^\top \bm{1}\{X_i = x\} + O(b) \\
     & = \Sigma_{n^\calI}(x) + O_P(1/\sqrt{n^\calI} + b).
    \end{split} 
\end{align}
Next, write
\begin{align}
    \sqrt{n^\calI} \left( \hat \beta(x) - \beta(x)\right) 
    & = \left( \frac{1}{n^\calI} \sum_{i \in \calI} W_i W_i^\top L_{i,b}(x) \right)^{-1} \frac{1}{\sqrt{n^\calI}} \sum_{i \in \calI} W_i (Y_i - W_i^\top \beta(x))L_{i,b}(x) \\
    & = A_1(x) + A_2(x) + A_3(x),
\end{align}
where
\begin{align}
    A_1(x)
    & \coloneqq \left( \frac{1}{n^\calI} \sum_{i \in \calI} W_i W_i^\top L_{i,b}(x) \right)^{-1} \frac{1}{\sqrt{n^\calI}} \sum_{i \in \calI}  W_i W_i^\top \left\{ \beta(X_i) - \beta(x)\right\} L_{i,b}(x) \\
    A_2(x) 
    & \coloneqq \left( \frac{1}{n^\calI} \sum_{i \in \calI} W_i W_i^\top L_{i,b}(x) \right)^{-1} \frac{1}{\sqrt{n^\calI}} \sum_{i \in \calI} W_i \epsilon_i \bm{1}\{X_i = x\} \\
    A_3(x) 
    & \coloneqq \left( \frac{1}{n^\calI} \sum_{i \in \calI} W_i W_i^\top L_{i,b}(x) \right)^{-1} \frac{1}{\sqrt{n^\calI}} \sum_{i \in \calI} W_i \epsilon_i \ell_{i,b}(x).
\end{align}
Observe that, for all $x \in \calX$,
\begin{align}
    A_1(x)
    & = \left( \frac{1}{n^\calI} \sum_{i \in \calI} W_i W_i^\top L_{i,b}(x) \right)^{-1} \frac{1}{\sqrt{n^\calI}} \sum_{i \in \calI}  W_i W_i^\top \left\{ \beta(X_i) - \beta(x)\right\} \bm{1}\{X_i = x\} \\
    & \quad + \left( \frac{1}{n^\calI} \sum_{i \in \calI} W_i W_i^\top L_{i,b}(x) \right)^{-1} \frac{1}{\sqrt{n^\calI}} \sum_{i \in \calI}  W_i W_i^\top \left\{ \beta(X_i) - \beta(x)\right\} \ell_{i,b}(x) \\
    & = O_P\left(\sqrt{n^\calI} b\right).
\end{align}
For $A_3(x)$,
\begin{align}
    \bbE \left\| \frac{1}{\sqrt{n^\calI}} \sum_{i \in \calI} W_i \epsilon_i \ell_{i,b}(x) \right\|^2 
    & = \frac{1}{n^\calI} \sum_{i, i' \in \calI} \bbE\left[ W_i^\top W_{i'} \epsilon_i \epsilon_{i'} \ell_{i,b}(x) \ell_{i',b}(x) \right] \\
    & = \frac{1}{n^\calI} \sum_{i, i', j, j' \in \calI} \bbE\left[ W_i^\top W_{i'} r_{ij} r_{i'j'} \varepsilon_j \varepsilon_{j'} \ell_{i,b}(x) \ell_{i',b}(x) \right]  \\
    & = \frac{ \sigma^2_\varepsilon}{n^\calI} \sum_{i, i', j \in \calI}  r_{ij} r_{i'j} \bbE\left[ W_i^\top W_{i'}  \ell_{i,b}(x) \ell_{i',b}(x) \right] \\
    & \le \frac{\sigma^2_\varepsilon c^2 b^2}{n^\calI} \sum_{i, i', j \in \calI} |r_{ij}| \cdot | r_{i'j}|  = O(b^2).
\end{align}
Hence, by \eqref{eq:Sigma} and Markov's inequality, we have $A_3(x) = O_P(b)$.
Hence, we have $\sqrt{n^\calI} ( \hat \beta(x) - \beta(x) ) = A_2(x) + o_P(1)$ by Assumption \ref{as:vc}.5.

To apply the central limit theorem to $A_2(x)$, define
\begin{align}
    a_j \coloneqq  \bm{c}^\top \left( \Omega_{n^\calI}(x) \right)^{-1/2} \frac{1}{\sqrt{n^\calI}} \sum_{i \in \calI}  W_i \bm{1}\{X_i = x\} r_{ij} \varepsilon_j 
\end{align}
where $\bm{c} \in \bbR^{d_w}$ satisfying $||\bm{c}|| = 1$.
Note that $\bbE[a_j] = 0$ and $\sum_{j \in \calI} \bbE[a_j^2] = 1$ hold:
\begin{align}
    \sum_{j \in \calI} \bbE[a_j^2] = \bm{c}^\top \left( \Omega_{n^\calI}(x) \right)^{-1/2} \frac{1}{n^\calI} \sum_{i, i',j  \in \calI} \bbE[ W_i  W_{i'}^\top \bm{1}\{X_i = x, X_{i'} = x\} r_{ij} r_{i'j}\varepsilon_j \varepsilon_j  ] \left( \Omega_{n^\calI}(x) \right)^{-1/2} \bm{c} = 1
\end{align}
by Assumption \ref{as:vc}.2.
Moreover, by Assumptions \ref{as:vc}.1, Assumptions \ref{as:vc}.2, and Assumptions \ref{as:vc}.3,
\small\begin{align}
    \sum_{j \in \calI} \bbE[a_j^4]
    & = \frac{1}{(n^\calI)^2} \sum_{j \in \calI} \sum_{i_1, i_2, i_3, i_4 \in \calI} \bbE\left[  r_{i_1 j} r_{i_2 j} W_{i_1}^\top \left( \Omega_{n^\calI}(x) \right)^{-1/2} \bm{c} \bm{c}^\top \left( \Omega_{n^\calI}(x) \right)^{-1/2} W_{i_2} \right. \\
    & \quad \times \left. r_{i_3 j} r_{i_4 j} W_{i_3}^\top \left( \Omega_{n^\calI}(x) \right)^{-1/2} \bm{c} \bm{c}^\top \left( \Omega_{n^\calI}(x) \right)^{-1/2} W_{i_4} \cdot \bm{1}\{X_{i_1} = X_{i_2} = X_{i_3} = X_{i_4} = x\}  \varepsilon_j^4 \right] \\
    & \le \frac{c}{(n^\calI)^2} \sum_{j \in \calI} \sum_{i_1, i_2, i_3, i_4 \in \calI} |r_{i_1 j}| \cdot |r_{i_2 j}| \cdot |r_{i_3 j}| \cdot |r_{i_4 j}| \bbE\left[ W_{i_1}^\top \left( \Omega_{n^\calI}(x) \right)^{-1} W_{i_2} W_{i_3}^\top \left( \Omega_{n^\calI}(x) \right)^{-1} W_{i_4} \right] \\
    & \le \frac{c c_w^4 c_r^4 }{n^\calI} \to 0.
\end{align}\normalsize
Then, by Lyapunov's central limit theorem, we obtain $\sum_{j \in \calI} a_j \overset{d}{\to} N(0,1)$.
Finally, by \eqref{eq:Sigma} and Slutsky's theorem,
\begin{align}
    \sqrt{n^\calI} \left( \hat \beta(x) - \beta(x) \right)  \overset{d}{\to} N\left( \bm{0}_{d_w}, (\Sigma_\calI(x))^{-1} \Omega_\calI(x) (\Sigma_\calI(x))^{-1} \right),
\end{align}
where $\Sigma_\calI(x) \coloneqq \lim_{n^\calI \to \infty} \Sigma_{n^\calI}(x)$, and $\Omega_\calI(x) \coloneqq \lim_{n^\calI \to \infty} \Omega_{n^\calI}(x)$.

\bigskip

(ii) By definition,
\begin{align}
    \sqrt{n^\calI} (\hat{\bm m} - \bm{m})
    & = \sqrt{n^\calI} \bm{Z} \left( \hat{\bm{\beta}} - \bm{\beta}\right) \\
    & = \bm{Z} \hat{\bm{J}}_{n^\calI} \frac{1}{\sqrt{n^\calI}} \sum_{i \in \calI}\left(
    \begin{array}{c}
     W_i \bm{1}\{X_i = x_1\} \\
     W_i \bm{1}\{X_i = x_2\} \\
    \vdots \\
     W_i \bm{1}\{X_i = x_{d_x}\}
    \end{array}
     \right) \epsilon_i + o_P(1),
\end{align}
where the last equality follows from the same argument as above.
Note that we have $\hat{\bm{J}}_{n^\calI} \overset{p}{\to} \bm{J}_{n^\calI}$ by Assumption \ref{as:vc}.4.
Similarly as above, we define
\begin{align}
    \bm{a}_j \coloneqq  \bm c^\top \left(\bm{\Omega}_{n^\calI}\right)^{-1/2} \frac{1}{\sqrt{n^\calI}} \sum_{i \in \calI}  \left(
    \begin{array}{c}
   W_i \bm{1}\{X_i = x_1\} \\
   W_i \bm{1}\{X_i = x_2\} \\
    \vdots \\
   W_i \bm{1}\{X_i = x_{d_x}\}
    \end{array}
     \right) r_{ij} \varepsilon_j, 
\end{align}
for any $\bm{c} \in \bbR^{d_w d_x}$ satisfying $||\bm{c}|| = 1$.
Then, by verifying the Lyapunov condition, we obtain $\sum_{j \in \calI}\bm{a}_j \overset{d}{\to} N(0,1)$, which implies the desired result:
\begin{align}
    \sqrt{n^\calI} (\hat{\bm m} - \bm{m}) \overset{d}{\to} N\left( \bm{0}_{d_x d_g}, \bm{Z} \bm{J}_\calI \bm{\Omega}_\calI \bm{J}_\calI \bm{Z}^\top \right)
\end{align}
by Slutsky's theorem, where $\bm{J}_\calI \coloneqq \lim_{n^\calI \to \infty} \bm{J}_{n^\calI}$, and $\bm{\Omega}_\calI  \coloneqq \lim_{n^\calI \to \infty} \bm{\Omega}_{n^\calI}$.

\qed


\begin{flushleft}
    \textbf{Proof of Theorem \ref{thm:dist}}
\end{flushleft}

Define
\begin{align}
    \phi(\bm m) \coloneqq \sum_{x \in \calX} \left[ \max_{\Gamma_x \in \calB_{\delta, q, x}} \sum_{u,v \in \calG^2} \Gamma_x(u,v) m(v, x)\right].
\end{align}
Then, we can write concisely $\bar \kappa_{\delta, q} = \phi(\bm{m})$ and $\hat{\bar \kappa}_{\delta, q} = \phi(\hat{\bm{m}})$.

By Theorem 2.1 of \cite{fang2019inference} (see also \cite{shapiro1991asymptotic}), we know that 
\begin{align}
    \sqrt{n^\calI}( \hat{\bar \kappa}_{\delta, q} - \bar \kappa_{\delta, q}) 
    & = \sqrt{n^\calI}( \phi(\hat{\bm m}) - \phi(\bm m) ) \\
    & = \phi_{\bm m}'(\sqrt{n^\calI} (\hat{\bm m} - \bm{m})) + o_P(1),
\end{align}
and therefore $\sqrt{n^\calI}( \hat{\bar \kappa}_{\delta, q} - \bar \kappa_{\delta, q}) \overset{d}{\to} \phi_{\bm m}'(N( \bm{0}_{d_x d_g}, \bm{Z} \bm{J}_\calI \bm{\Omega}_\calI \bm{J}_\calI \bm{Z}^\top ))$, where $\phi_{\bm m}'(\bm h)$ is the Hadamard directional derivative of $\phi$ at $\bm m$ in the direction $\bm h \in \bbR^{d_g d_x}$.

The explicit form of $\phi'$ can be derived as follows.
Let us denote
\begin{align}
    & \langle \Gamma, f \rangle \coloneqq \sum_{u,v \in \calG^2} \Gamma(u,v) f(v)\\
    & \phi(x, f) \coloneqq \max_{\Gamma \in \calB_{\delta, q, x}} \langle \Gamma, f \rangle,
\end{align}
so that $\phi(\bm m) = \sum_{x \in \calX} \phi(x, m(\cdot, x))$.
Define 
\begin{align}
    \calS^*_{\delta, q, x}(f)
    & \coloneqq \argmax_{\Gamma \in \calB_{\delta, q, x}} \: \langle \Gamma, f \rangle.
\end{align}
Consider any sequence $h_t \to h \in \bbR^{d_g}$ as $t \downarrow 0$.
Observe that, for $\Gamma_t \in \calS^*_{\delta, q, x}(f + t h_t)$ and $\Gamma_0 \in \calS^*_{\delta, q, x}(f)$,
\begin{align}
    \frac{\phi(x, f + t h_t) - \phi(x, f)}{t} = \frac{\langle \Gamma_t, f + t h_t \rangle - \langle \Gamma_0, f \rangle}{t} 
    & = \frac{\langle \Gamma_t, f \rangle + t \langle \Gamma_t, h_t \rangle - \langle \Gamma_0, f \rangle}{t}  \\
    & \le \langle \Gamma_t, h_t \rangle,
\end{align}
where the last inequality follows because $\langle \Gamma_t, f \rangle \le \langle \Gamma_0, f \rangle$.
Since $\Gamma_t$ is a sequence in $\calB_{\delta, q, x}$ and $\calB_{\delta, q, x}$ is compact, the right-hand side converges to $\langle \Gamma_0, h \rangle$,
leading to 
\begin{align}\label{eq:upper}
    \lim\sup_{t \downarrow 0}\frac{\langle \Gamma_t, f + t h_t \rangle - \langle \Gamma_0, f \rangle}{t} \le \max_{\Gamma \in \calS^*_{\delta, q, x}(f)} \langle \Gamma, h \rangle.
\end{align}
Meanwhile,
\begin{align}
    \frac{\langle \Gamma_t, f + t h_t \rangle}{t} & \ge \frac{\langle \Gamma_0, f + t h_t \rangle}{t} \\
    & = \frac{\langle \Gamma_0, f \rangle}{t} + \langle \Gamma_0, h_t \rangle.
\end{align}
Since the above holds for all $\Gamma_0 \in \calS^*_{\delta, q, x}(f)$, 
\begin{align}\label{eq:lower}
    \lim\inf_{t \downarrow 0}\frac{\langle \Gamma_t, f + t h_t \rangle - \langle \Gamma_0, f \rangle}{t} \ge \max_{\Gamma \in \calS^*_{\delta, q, x}(f)} \langle \Gamma, h \rangle.
\end{align}
From \eqref{eq:upper} and \eqref{eq:lower}, we can find that $\phi'_f(x, h) = \max_{\Gamma \in \calS^*_{\delta, q, x}} \langle \Gamma, h \rangle$. 
 
Hence, in our context, writing $\calS^*_{\delta, q, x} \coloneqq \argmax_{\Gamma \in \calB_{\delta, q, x}} \: \langle \Gamma, m(\cdot, x) \rangle$,
\begin{align}
    \phi'_{\bm m}(\bm h) 
    & = \lim_{t \downarrow 0} \frac{\sum_{x \in \calX} \phi(x, m(\cdot, x) + t h_t(\cdot, x)) - \phi(x, m(\cdot, x))}{t} \\
    & = \sum_{x \in \calX} \left[ \max_{\Gamma_x \in \calS^*_{\delta, q, x}} \sum_{u,v \in \calG^2} \Gamma_x(u,v) h(v, x) \right].
\end{align}
Consequently,
\begin{align}
    \sqrt{n^\calI}\left( \hat{\bar \kappa}_{\delta, q} - \bar \kappa_{\delta, q}\right) 
    & = \phi_{\bm m}'(\sqrt{n^\calI} (\hat{\bm m} - \bm{m})) + o_P(1) \\
    & = \sum_{x \in \calX} \left[ \max_{\Gamma_x \in \calS_{\delta, q, x}^*} \sqrt{n^\calI} \sum_{u,v \in \calG^2} \Gamma_x(u,v) \left(\hat{m}(v, x) - m(v, x) \right) \right] + o_P(1)\\
    & \overset{d}{\to} \sum_{x \in \calX} \left[ \max_{\Gamma_x \in \calS_{\delta, q, x}^*} \sum_{u,v \in \calG^2} \Gamma_x(u,v) \bbG(v, x) \right]
\end{align}
by Proposition \ref{prop:m}(ii).

\qed


\begin{flushleft}
    \textbf{Proof of Theorem \ref{thm:boot}}
\end{flushleft}
Let $\epsilon^*_i \coloneqq \eta_i \hat \epsilon_i$.
Write
\begin{align}
    \sqrt{n^\calI} \left( \hat \beta^*(x) - \hat \beta(x)\right) 
    & = \left( \frac{1}{n^\calI} \sum_{i \in \calI} W_i W_i^\top L_{i,b}(x) \right)^{-1} \frac{1}{\sqrt{n^\calI}} \sum_{i \in \calI} W_i Y_i^* L_{i,b}(x) -  \sqrt{n^\calI} \hat \beta(x) \\
    & = A_1^*(x) + A_2^*(x) + A_3^*(x),
\end{align}
where
\begin{align}
    A_1^*(x)
    & \coloneqq \left( \frac{1}{n^\calI} \sum_{i \in \calI} W_i W_i^\top L_{i,b}(x) \right)^{-1} \frac{1}{\sqrt{n^\calI}} \sum_{i \in \calI}  W_i W_i^\top \left\{ \hat \beta(X_i) - \hat \beta(x)\right\} L_{i,b}(x) \\
    A_2^*(x) 
    & \coloneqq \left( \frac{1}{n^\calI} \sum_{i \in \calI} W_i W_i^\top L_{i,b}(x) \right)^{-1} \frac{1}{\sqrt{n^\calI}} \sum_{i \in \calI} W_i \epsilon_i^* \bm{1}\{X_i = x\} \\
    A_3^*(x) 
    & \coloneqq \left( \frac{1}{n^\calI} \sum_{i \in \calI} W_i W_i^\top L_{i,b}(x) \right)^{-1} \frac{1}{\sqrt{n^\calI}} \sum_{i \in \calI} W_i \epsilon_i^* \ell_{i,b}(x).
\end{align}
Analogously to the proof of Proposition \ref{prop:m}, we can easily find that $A_1^*(x) = O_P\left(\sqrt{n^\calI} b\right)$.
For $A_3^*(x)$, decompose $A_3^*(x) = A_{31}^*(x) + A_{32}^*(x)$, where
\begin{align}
    A_{31}^*(x) 
    & \coloneqq \left( \frac{1}{n^\calI} \sum_{i \in \calI} W_i W_i^\top L_{i,b}(x) \right)^{-1} \frac{1}{\sqrt{n^\calI}} \sum_{i \in \calI} W_i (\epsilon_i^* - \eta_i \epsilon_i) \ell_{i,b}(x) \\
    A_{32}^*(x) 
    & \coloneqq \left( \frac{1}{n^\calI} \sum_{i \in \calI} W_i W_i^\top L_{i,b}(x) \right)^{-1} \frac{1}{\sqrt{n^\calI}} \sum_{i \in \calI} W_i \eta_i \epsilon_i \ell_{i,b}(x).
\end{align}
Noting that $\hat \epsilon_i - \epsilon_i = Y_i - W_i^\top \hat \beta(X_i) - \epsilon_i = W_i^\top (\beta(X_i) - \hat \beta(X_i))$, write
\begin{align}
    W_i (\epsilon_i^* - \eta_i \epsilon_i) \ell_{i,b}(x)
    & = W_i \eta_i (\hat \epsilon_i - \epsilon_i)\ell_{i,b}(x) \\
    & = W_i W_i^\top (\beta(X_i) - \hat \beta(X_i)) \eta_i \ell_{i,b}(x) \\
    & \eqqcolon c_i \eta_i \ell_{i,b}(x),
\end{align}
where $c_i = O_P(1/\sqrt{n^\calI})$ by Proposition \ref{prop:m}(i).
Further,
\begin{align}
    \bbE^* \left\| \frac{1}{\sqrt{n^\calI}} \sum_{i \in \calI} W_i (\epsilon_i^* - \eta_i \epsilon_i) \ell_{i,b}(x) \right\|^2 
    & =  \frac{1}{n^\calI} \sum_{i, i' \in \calI} \bbE^*\left[ c_i^\top c_{i'} \eta_i \eta_{i'} \ell_{i,b}(x) \ell_{i',b}(x)\right] \\
    & = \frac{1}{n^\calI} \sum_{i, i' \in \calI} \bbE^*[ \eta_i \eta_{i'} ] c_i^\top c_{i'} \ell_{i,b}(x) \ell_{i',b}(x).
\end{align}
Recall that $\bbE^*[\bm \eta \bm \eta^\top] = \bbK_\calI$, and hence $\bbE^*[ \eta_i \eta_{i'} ] = K(\tilde \Delta_{i, i'}/d)$.
Then,
\begin{align}
    \frac{1}{n^\calI} \sum_{i, i' \in \calI} c_i^\top c_{i'} \ell_{i,b}(x) \ell_{i',b}(x) K\left( \frac{\tilde \Delta_{i, i'}}{d} \right)
    & \le O_P(b^2) \frac{1}{(n^\calI)^2} \sum_{i, i' \in \calI} \left| K\left( \frac{\tilde \Delta_{i, i'}}{d} \right) \right| \\
    & = o_P(b^2),  
\end{align}
where the last line is due to Lemma A.1 of \cite{conley2023bootstrap}.
Hence, by Assumption \ref{as:vc}.5 and Markov's inequality with \eqref{eq:Sigma}, we have $A_{31}^*(x) = o_{P^*}(1)$ in probability.
Similarly,
\begin{align}
    \bbE \left(\bbE^* \left\| \frac{1}{\sqrt{n^\calI}} \sum_{i \in \calI} W_i \eta_i \epsilon_i \ell_{i,b}(x) \right\|^2 \right)
    & =  \frac{1}{n^\calI} \sum_{i, i' \in \calI} \bbE \left[ W_i^\top W_i \epsilon_{i} \epsilon_{i'} \bbE^*[ \eta_i \eta_{i'} ] \ell_{i,b}(x) \ell_{i',b}(x)\right] \\
    & = \frac{1}{n^\calI} \sum_{i, i' \in \calI} \bbE \left[ W_i^\top W_i \epsilon_{i} \epsilon_{i'} \ell_{i,b}(x) \ell_{i',b}(x) K\left( \frac{\tilde \Delta_{i, i'}}{d} \right) \right] = O(b^2),  
\end{align}
implying that $A_{32}^*(x) = o_{P^*}(1)$ in probability.

We apply the same decomposition to $A_2^*(x)$: $A_2^*(x) = A_{21}^*(x) + A_{22}^*(x)$,
\begin{align}
    A_{21}^*(x) 
    & \coloneqq \left( \frac{1}{n^\calI} \sum_{i \in \calI} W_i W_i^\top L_{i,b}(x) \right)^{-1} \frac{1}{\sqrt{n^\calI}} \sum_{i \in \calI} W_i (\epsilon_i^* - \eta_i \epsilon_i) \bm{1}\{X_i = x\} \\
    A_{22}^*(x) 
    & \coloneqq \left( \frac{1}{n^\calI} \sum_{i \in \calI} W_i W_i^\top L_{i,b}(x) \right)^{-1} \frac{1}{\sqrt{n^\calI}} \sum_{i \in \calI} W_i \eta_i \epsilon_i \bm{1}\{X_i = x\}.
\end{align}
Then, by the same argument as in the evaluation of $A^*_{31}(x)$, it is straightforward to see that $A_{21}^*(x) = o_{P^*}(1)$ in probability.

Since the above discussion applies to all $x \in \calX$, consequently, we have 
\begin{align}
    \sqrt{n^\calI} (\hat{\bm m}^* - \hat{\bm m})
    & = \sqrt{n^\calI} \bm{Z} \left( \hat{\bm \beta}^* - \hat{\bm \beta}\right) \\
    & = \bm{Z} \hat{\bm{J}}_{n^\calI} \frac{1}{\sqrt{n^\calI}} \sum_{i \in \calI}\left(
    \begin{array}{c}
     W_i \bm{1}\{X_i = x_1\} \\
     W_i \bm{1}\{X_i = x_2\} \\
    \vdots \\
     W_i \bm{1}\{X_i = x_{d_x}\}
    \end{array}
     \right) \eta_i \epsilon_i + o_{P^*}(1),
\end{align}
with probability approaching one, where the definitions of $\hat{\bm m}^*$ and $\hat{\bm \beta}^*$ should be clear from the context.
Furthermore, following the same argument as in the proof of Theorem 3.1 (equation (20)) of \cite{conley2023bootstrap}, we obtain
\begin{align}
     \left(\bm{\Omega}_{n^\calI}\right)^{-1/2}  \frac{1}{\sqrt{n^\calI}} \sum_{i \in \calI}\left(
    \begin{array}{c}
     W_i \bm{1}\{X_i = x_1\} \\
     W_i \bm{1}\{X_i = x_2\} \\
    \vdots \\
     W_i \bm{1}\{X_i = x_{d_x}\}
    \end{array}
     \right) \eta_i \epsilon_i \overset{d^*}{\to} N(\bm 0_{d_w d_x}, I_{d_w d_x})
\end{align}
in probability, and hence
\begin{align}\label{eq:boot_m}
    {\Pr}^*\left(\sqrt{n^\calI} (\hat{\bm m}^* - \hat{\bm m}) \le s\right) = \Pr\left( \sqrt{n^\calI} (\hat{\bm m} - \bm m) \le s\right) + o_P(1)
\end{align}
uniformly in $s \in \bbR$.
Then, in view of Proposition \ref{prop:m}(ii) and Theorem \ref{thm:dist}, we can see that $\sqrt{n^\calI}(\hat{\bar \kappa}^*_{\delta, q} - \hat{\bar \kappa}_{\delta, q} )$ and $\sqrt{n^\calI}( \hat{\bar \kappa}_{\delta, q} - \bar \kappa_{\delta, q})$ share the same asymptotic distribution conditional on the event $\{\hat \calS_{\delta, q, x}^* = \calS_{\delta, q, x}^*\}$.

In view of the proof of Theorem \ref{thm:dist}, we can see that $\hat{\bar{\kappa}}_{\delta, q, x} = \bar{\kappa}_{\delta, q, x} + O_P(1/\sqrt{n^\calI})$, where $\bar{\kappa}_{\delta, q, x} \coloneqq \max_{\Gamma \in \calB_{\delta, q, x}} \sum_{u, v \in \calG^2} \Gamma(u, v) m(v, x)$.
Suppose that $\Gamma \in \calS_{\delta, q, x}^*$.
Then,
\begin{align}
\sum_{u,v \in \calG^2} \Gamma(u,v) \hat{m}(v, x) 
    & =  \sum_{u,v \in \calG^2} \Gamma(u,v) m(v, x) + \sum_{u,v \in \calG^2} \Gamma(u,v) (\hat m(v, x) - m(v,x)) \\
    & = \bar{\kappa}_{\delta, q, x} + O_P(1/\sqrt{n^\calI}) \\
    & = \hat{\bar{\kappa}}_{\delta, q, x} + O_P(1/\sqrt{n^\calI}) \\
    & \ge \hat{\bar{\kappa}}_{\delta, q, x} - a
\end{align}
with probability approaching one under Assumption \ref{as:boot}.6.
This implies that $\Pr(\calS_{\delta, q, x}^* \subseteq \hat{\calS}_{\delta, q, x}^*) \to 1$ as $n^\calI \to \infty$. 
On the other hand, suppose that $\Gamma \in \hat{\calS}_{\delta, q, x}^*$.
Then,
\begin{align}
    \sum_{u,v \in \calG^2} \Gamma(u,v) m(v, x) 
    & =  \sum_{u,v \in \calG^2} \Gamma(u,v) \hat m(v, x) + \sum_{u,v \in \calG^2} \Gamma(u,v) ( m(v, x) - \hat m(v,x)) \\
    &  \ge \hat{\bar{\kappa}}_{\delta, q, x} - a + O_P(1/\sqrt{n^\calI}) \\
    &  = \bar{\kappa}_{\delta, q, x} - a + O_P(1/\sqrt{n^\calI}).
\end{align}
Here, note that if $\Gamma \notin \calS_{\delta, q, x}^*$, then the strict inequality $\sum_{u,v \in \calG^2} \Gamma(u,v) m(v, x) < \bar{\kappa}_{\delta, q, x}$ must hold.
Hence, since $a + O_P(1/\sqrt{n^\calI})$ converges to zero in probability as $n^\calI$ increases, the above inequality implies that $\Gamma \in \calS_{\delta, q, x}^*$ holds with probability approaching one; that is, $\Pr(\hat{\calS}_{\delta, q, x}^* \subseteq \calS_{\delta, q, x}^*) \to 1$.
Hence, $\Pr(\calS_{\delta, q, x}^* = \hat{\calS}_{\delta, q, x}^*) \to 1$.

\qed

\section{Estimation and Bootstrap Inference for the Lower Bound}\label{app:lower}

The estimation of the lower bound $\underline{\kappa}_{\delta, q}$ can be performed by solving the following linear programming:
\begin{align}\begin{split}
    & \hat{\underline{\kappa}}_{\delta, q} \coloneqq  \sum_{x \in \calX} \left[ \min_{\Gamma_x} \sum_{u,v \in \calG^2} \Gamma_x(u,v) \hat m(v, x) \right]\\
    &\text{subject to } \sum_{v \in \calG}\Gamma_x(u, v) = \pi_x^*(u), \sum_{u, v \in \calG^2} \Gamma_x(u, v) \bigl|u - v\bigr|^q \le \delta^q, \Gamma_x(u, v) \ge 0, \forall\, (x, u, v) \in \calX \times \calG^2,
\end{split}\end{align}
where $\hat m$ is obtained through the varying-coefficient estimation as in Subsection \ref{subsec:vcoef}.

To describe the wild bootstrap procedure for the lower bound, let 
\begin{align}
    \hat{\calT}^*_{\delta, q, x} \coloneqq 
    \left\{ \Gamma \in \calB_{\delta, q, x} : 
    \sum_{u,v \in \calG^2} \Gamma(u,v) \hat{m}(v, x) 
    \le \hat{\underline{\kappa}}_{\delta, q, x} + a \right\},
\end{align}
which is considered as the estimator of $\calT^*_{\delta, q, x} \coloneqq \argmin_{\Gamma \in \calB_{\delta, q, x}} \sum_{u, v \in \calG^2} \Gamma(u, v) m(v, x)$.
Then, the distribution of $\sqrt{n^\calI}(\hat{\underline{\kappa}}_{\delta, q} - \underline{\kappa}_{\delta, q})$ can be simulated in the following manner.

\begin{algorithm}[H]
\caption{Wild bootstrap procedure for inference on $\underline{\kappa}_{\delta,q}$}
\begin{algorithmic}[1]
\State Estimate $\hat \beta(x)$ for all $x \in \calX$ using \eqref{eq:coef}
\State Compute the residual $\hat \epsilon_i \coloneqq Y_i - W_i^\top \hat \beta(X_i)$ for all $i \in \calI$
\For{$b = 1$ to $B$}
    \State Draw $\bm{\eta}^{(b)} = (\eta_1^{(b)}, \ldots, \eta_{n^{\calI}}^{(b)}) \sim \Phi_\calI \Lambda_\calI^{1/2} N(\bm{0}_{n^\calI}, I_{n^\calI})$
    \State Generate a bootstrap sample $\{(W_i, Y_i^{*(b)}): i \in \calI\}$, where $Y_i^{*(b)} \coloneqq W_i^\top \hat \beta(X_i) + \eta_i^{(b)} \hat \epsilon_i$
    \State Obtain $\hat \beta^{*(b)}(x)$ by the kernel weighted regression of $Y_i^{*(b)}$ on $W_i$ for all $x \in \calX$
    \State Compute $\hat{\underline \kappa}_{\delta,q}^{*(b)} \coloneqq \sqrt{n^{\calI}} \sum_{x \in \calX}  \left[ \min_{\Gamma_x \in \hat{\calT}_{\delta, q, x}^*} \sum_{u, v \in \calG^2} \Gamma_x(u,v) z(v, x)^\top (\hat \beta^{*(b)}(x) - \hat \beta(x)) \right]$
\EndFor
\State Compute the empirical $\alpha$ quantile $\hat \omega_{B, \alpha/2}$ of $\left\{\sqrt{n^\calI}(\hat{\underline \kappa}_{\delta,q}^{*(b)} - \hat{\underline \kappa}_{\delta, q}) : b = 1, \ldots, B\right\}$
\end{algorithmic}
\end{algorithm}
Further, the asymptotic $100(1 - \alpha)$\% CI for $\underline \kappa_{\delta, q}$ can be obtained by
\begin{align}
    \calC_{1 - \alpha}(\underline \kappa_{\delta, q})
    \coloneqq \left[\hat{\underline \kappa}_{\delta, q} - \frac{\hat \omega_{B, 1 - \alpha/2}}{\sqrt{n^\calI}}, \:  \hat{\underline \kappa}_{\delta, q} - \frac{\hat \omega_{B, \alpha/2}}{\sqrt{n^\calI}} \right] .
\end{align}

\clearpage
\bibliography{references}

\end{document}